\documentclass[12pt]{iopart}
\usepackage{iopams}
\expandafter\let\csname equation*\endcsname\relax
\expandafter\let\csname endequation*\endcsname\relax
\usepackage{amsmath}
\usepackage{physics} 
\usepackage{tikz}
\usepackage{comment}

\usepackage{graphicx}
\usepackage{xcolor}
\usepackage{caption,subcaption}

\usepackage[colorlinks=true]{hyperref}
\hypersetup{
    linktoc    = all,
    colorlinks = true,
    citecolor  = blue,
    urlcolor   = magenta,
    linkcolor  = magenta
}
\begin{document}

\paper[Dissipation alters modes of information encoding in small quantum reservoirs]{Dissipation alters modes of information encoding in small quantum reservoirs near criticality}

\author{Krai Cheamsawat\textsuperscript{1} and Thiparat Chotibut\textsuperscript{1}\footnote{Correspondence to: \mailto{thiparatc@gmail.com}}}
\address{\textsuperscript{1}Chula Intelligent and Complex Systems Lab, Department of Physics, Faculty of Science, Chulalongkorn University, Bangkok, Thailand, 10330}
\eads{\mailto{kraicwt@gmail.com}, \mailto{thiparatc@gmail.com}}

\vspace{6pt}
\begin{indented}
\item[]\today
\end{indented}

\begin{abstract}
Quantum reservoir computing (QRC) has emerged as a promising paradigm for harnessing near-term quantum devices to tackle temporal machine learning tasks.  Yet identifying the mechanisms that underlie enhanced performance remains challenging, particularly in many-body open systems where nonlinear interactions and dissipation intertwine in complex ways. Here, we investigate a minimal model of a driven-dissipative quantum reservoir described by two coupled Kerr-nonlinear oscillators, an experimentally realizable platform that features controllable coupling, intrinsic nonlinearity, and tunable photon loss. Using Partial Information Decomposition (PID), we examine how different dynamical regimes encode input drive signals in terms of {\it redundancy} (information shared by each oscillator) and {\it synergy} (information accessible only through their joint observation). Our key results show that, near a critical point marking a dynamical bifurcation, the system transitions from predominantly redundant to synergistic encoding. We further demonstrate that synergy amplifies short-term responsiveness, thereby enhancing immediate memory retention, whereas strong dissipation leads to more redundant encoding that supports long-term memory retention. These findings elucidate how the interplay of instability and dissipation shapes information processing in small quantum systems, providing a fine-grained, information-theoretic perspective for analyzing and designing QRC platforms.
\end{abstract}


\section{Motivation and Introduction}

Reservoir Computing (RC) is a computational paradigm that harnesses the intrinsic dynamics of complex systems to process time-dependent inputs efficiently \cite{jaeger2001short, lukovsevivcius2009reservoir,nakajima2018reservoir}. Unlike conventional recurrent neural networks (RNNs), RC requires training only at the readout layer, circumventing expensive weight-update procedures on internal nodes \cite{lukovsevivcius2012practical}. Quantum Reservoir Computing (QRC) extends these ideas to quantum platforms, leveraging quantum superposition and entanglement to amplify the dimensionality of the feature space and potentially enhance computational capabilities \cite{fujii2017harnessing, nakajima2018reservoir}. Early demonstrations of QRC have shown promise in tasks like time-series prediction, classification, and memory capacity estimation, and ongoing efforts explore a range of theoretical and experimental strategies for improving performance \cite{Innocenti2023, motamedi2023correlations, gotting2023exploring, rodrigo_QRCfinite_PRE, sannia2024dissipation,  xiong2024QELM, Mujal2023, govia2021quantum, ivaki2024, sornsaeng2024}.

Recent QRC research has primarily focused on many-body quantum systems, where quantum phase transitions are suspected to boost computational expressivity \cite{martinez2021dynamical,Mujal_Zambrini_reviewQRC}. While numerical studies reveal intriguing heuristics, such as enhanced memory capacity near critical points, designing optimal quantum reservoirs remains an open question, partly due to the complexity of analyzing large quantum systems. Here, we adopt a complementary approach by studying a \emph{pair of coupled Kerr-nonlinear oscillators}, a minimal yet experimentally realizable quantum platform \cite{hartmann2008quantum}. This system exhibits rich dynamical behaviors, including dynamical instability (bifurcation) and dissipation due to photon loss that can be precisely tuned. As such, it offers a tractable yet nontrivial testbed for exploring how quantum correlation, instability, and dissipation govern quantum information processing.

To dissect how these coupled oscillators encode incoming signals, we draw on information-theoretic concepts from neuroscience, where measures of synergy and redundancy helped analyze how neural networks collectively encode stimuli \cite{gat1998synergy, brenner2000synergy, schneidman2003synergy, latham2005synergy}. As traditional mutual information metric fails to separate out redundant and synergistic contributions to information encoding, we employ \emph{Partial Information Decomposition} (PID) \cite{bertschinger2014quantifying}, which partitions the total information into three components: \emph{redundancy}, capturing information that both oscillators share; \emph{unique information}, provided by each oscillator individually; and \emph{synergy}, arising only when both oscillators are observed together. This perspective provides a fine-grained view into the internal encoding structure of the reservoir.

To connect the system’s dynamics to its information-encoding strategy, we combine numerical simulations with non-equilibrium mean-field theory based on the Keldysh formalism \cite{kamenev2023field}, focusing on how small external perturbations propagate through the system. In particular, we study how coupling strength, frequency detuning, and photon loss rate influence the system’s response, and then show how these distinct dynamical regimes lead to different synergy and redundancy profiles in the oscillators’ outputs.

Our main findings reveal that near a critical coupling strength leading to dynamical bifurcation, the system transitions from predominantly redundant encoding to a regime featuring significant synergistic information. This synergistic behavior arises from the interplay between fast {\it collective} oscillations and overdamped {\it soft} modes. We show that increasing dissipation suppresses quantum correlations, and promotes highly redundant encoding modes. In contrast, near the onset of dynamical instability, synergy is amplified and enriches short-term responsiveness, improving short-term memory retention. Taken together, these results highlight how dissipation and dynamic instability in a minimal system can steer a quantum reservoir toward redundant or synergistic processing, each regime benefiting different computational tasks. 

In relation to recent proposals using two coupled Kerr-nonlinear oscillators as quantum reservoirs \cite{Dudas2023, DudasIEEE2022}, which highlight the roles of dissipation for fading memory and moderate coupling for richer dynamics, our work differs by focusing on PID to examine how synergy and redundancy influences the reservoir’s memory capacity. This complements prior findings to show that critical points in Kerr dynamics can shift encoding from redundant to synergistic regimes. Meanwhile, single Kerr oscillators with large Hilbert spaces \cite{govia2021quantum, kalfus_PRR_2022} highlight how dimensionality alone can serve as a computational resource, but our key question of \emph{whether the whole can exceed the sum of its parts} requires at least two coupled oscillators for emergent synergistic encoding. Lastly, although \cite{khan2021} studies larger arrays of Kerr oscillators and demonstrates near-bifurcation enhancements of nonlinear memory and employ higher-order cumulant expansions to handle higher photon numbers, we limit ourselves to at most a second-order expansion in a parameter regime where it remains accurate, focusing on PID-based insights rather than reservoir benchmarks and thus complementing other findings in the literature.

To guide the reader, this paper is organized as follows. Section \ref{sec:model} introduces the coupled Kerr-oscillator model and the relevant information-theoretic measures, including PID and quantum mutual information. Section \ref{sec:results} then presents our core numerical findings on synergy and redundancy, comparing fully quantum dynamics with both mean-field and cumulant expansions analyses. We also discuss the mechanisms driving redundant and synergistic encoding and examine how dissipation influences these encoding modes. In Section \ref{sec:MC}, we connect these insights to the quantum reservoir’s memory capacity. Finally, we conclude in Section \ref{sec:conclusion} by summarizing our results and outlining directions for future research. A pedagogical overview of PID can be found in \ref{sec:PID} and \ref{app:PID}, and the details of Keldysh approach to linear response analysis is provided in \ref{app:Keldysh}.

\section{Quantum Model, Relevant Information Measures, and Performance Metrics} \label{sec:model}

We first introduce our quantum system to study the interplay of instability and dissipation and their influence on modes of information encoding. Then we outline  the key goals of our study, and introduce information measures and metrics to characterize our quantum systems, and finally describe the numerical methods to study them.

\subsection{Model}
We consider a minimal model that can exhibit dynamical transitions from simple to more complex dynamics, and also support both redundant and synergistic modes of information encoding: a pair of coupled Kerr-nonlinear oscillators. Such systems are well-studied in cavity quantum electrodynamics and nonlinear optics, where the interplay of nonlinearities and external driving yields rich dynamical behavior \cite{milburn1991quantum,milburn1997quantum,carusotto2013quantum,goto2021chaos}. By focusing on just two coupled cavities, each supporting a single mode at a particular resonance frequency, we avoid the complexity of many-body phases that arise in the thermodynamic limit, thus isolating the essential ingredients needed to explore the onset of coordinated information encoding behaviors in quantum systems. 

We work in a frame rotating at the driving frequency $\omega_F$, and the corresponding Hamiltonian is given by
\begin{equation}
\hat{H}(t) = J(\hat{a}_1^{\dagger}\hat{a}_2+\hat{a}_2^{\dagger}\hat{a}_1) + \sum_{i=1,2} \left(\Delta_i\hat{a}_i^{\dagger}\hat{a}_i  + \frac{1}{2}U_i\hat{a}_i^{\dagger 2}\hat{a}_i^2 + F(t)(\hat{a}_i^{\dagger}+\hat{a}_i)\right),
\label{eqn:Hamiltonian}
\end{equation}
 where $\hat{a}_i$ and $\hat{a}_i^\dagger$ are the annihilation and creation operators for the $i^\text{th}$ cavity mode, respectively. The parameter $J$ governs the coupling strength between the two cavities, enabling photon tunneling and collective mode formation \cite{milburn1997quantum}. The detuning $\Delta_i = \omega_i - \omega_F$ measures the offset of the $i^\text{th}$ cavity’s resonance frequency $\omega_i$ from the driving frequency. The nonlinear Kerr coefficient $U_i$ characterizes the anharmonicity of each mode which introduces photon-photon interactions essential for generating nonclassical states. And $F(t)$ is a common external drive that contains time-dependent information applied to both cavities. 
 
To observe transitions between redundant and synergistic information encoding, the two cavities must differ in their nonlinear properties. In particular, having distinct Kerr coefficients (\(U_1 \neq U_2\)) breaks symmetry and enables nontrivial interactions between the modes.  With these minimal ingredients (a coherent drive, a tunable coupling, and carefully chosen nonlinearities), this system provides a controlled setting to study the fundamental mechanisms underlying both redundant and synergistic coding in interacting quantum systems.
 
\begin{figure}[!ht]
\begin{center}
\includegraphics[keepaspectratio, width = 0.5\textwidth]{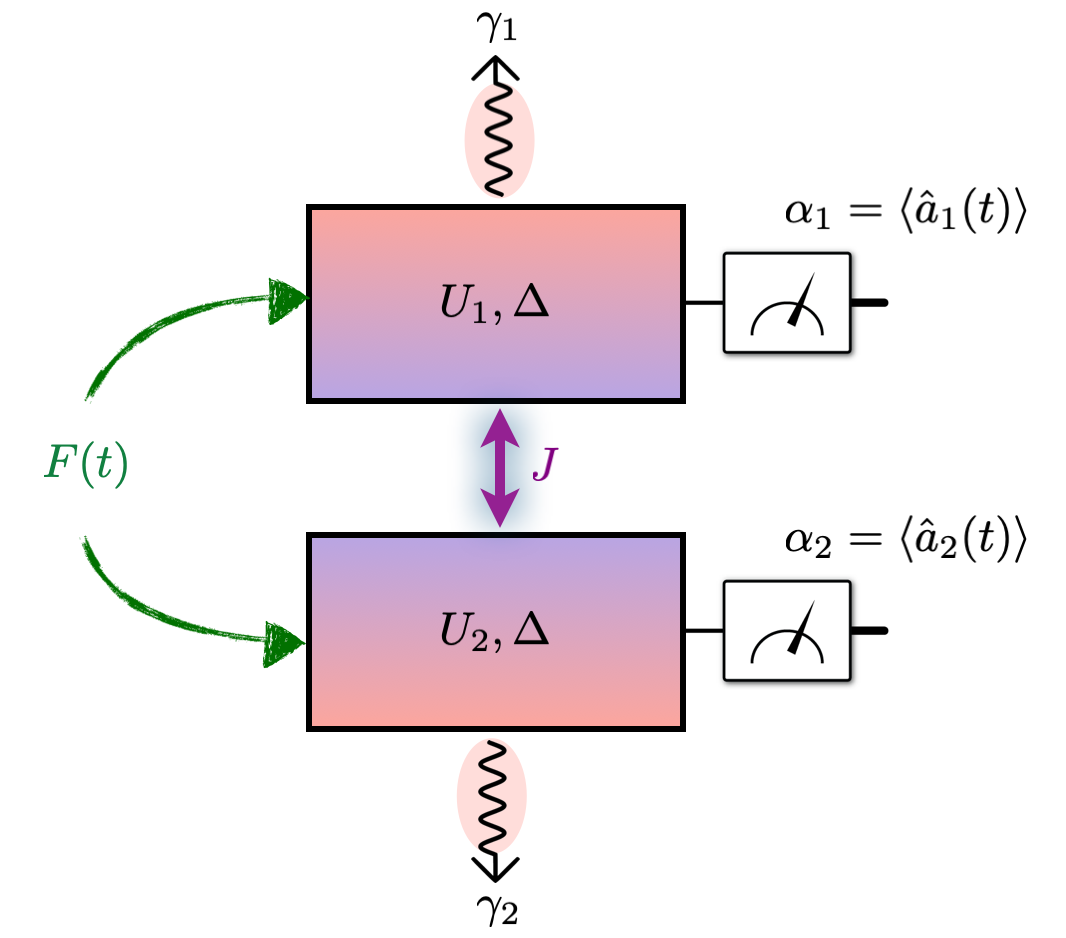}
\caption{Schematic of two coupled Kerr-nonlinear oscillators.
Each cavity $i$ features a Kerr nonlinearity $U_i$ and a photon-loss rate $\gamma_i$. A time-dependent drive $F(t)$ (green arrows) injects identical signals into both cavities, while coherent tunneling of strength $J$ (violet arrow) couples the two modes. We measure the mean fields $\alpha_i = \langle \hat{a}_i(t)\rangle$  to probe the system’s response. The Hamiltonian is specified by Eq.~\eqref{eqn:Hamiltonian}, and the Lindblad equation \eqref{eqn:Lindblad} governs this driven-dissipative dynamics. We assume both cavities have the same detuning $\Delta$ from the drive frequency. This work investigates how the readouts $\alpha_i(t)$ encode the time-dependent drive $F(t)$ across different dynamical regimes of the coupled Kerr oscillators.}
\label{fig:oscillator_schematic}
\end{center}
\end{figure}

To incorporate noise and dissipation, we consider the time evolution of an open quantum system weakly coupled to a Markovian bath. Specifically, we model the dynamics of the system’s density matrix \(\hat{\rho}(t)\) using the Lindblad master equation \cite{breuer2002theory}:
\begin{equation}    
\frac{d}{dt}\hat{\rho}(t) =  \mathcal{L}\hat{\rho}(t) = -i[\hat{H}(t), \hat{\rho}(t)] + \sum_{i =1,2}2\gamma_i\,\mathcal{D}[\hat{a}_i]\hat{\rho}(t),
\label{eqn:Lindblad}
\end{equation}
where \(\gamma_i\) is the photon decay rate of the \(i^\text{th}\) cavity associated with the Lindblad superoperator describing single-photon loss, \(\mathcal{D}[\hat{a}_i]\), which acts on the density matrix as
\begin{equation}
\mathcal{D}[\hat{a}_i]\hat{\rho} = \hat{a}_i \hat{\rho} \hat{a}_i^{\dagger} - \tfrac{1}{2}\{\hat{a}_i^{\dagger}\hat{a}_i, \hat{\rho}\}.
\end{equation}
In our simulations, we take \(\gamma_1 = \gamma_2 = \gamma\) for simplicity.

For the common time-dependent external driving field \(F(t)\), we choose \(F(t) = s(t)F\), where \(s(t)\) is a dimensionless, time-dependent signal, and \(F\) is a characteristic strength of the drive. To ensure that the system’s intrinsic dynamics dominate, we select \(F\) to be small  or comparable to other energy scales, and regard $F(t)$ as a perturbation. 
In this work, \(s(t)\) is taken to be a symmetric telegraph process with \(s(t) \in \{-1,1\}\) \cite{bena2006dichotomous}. While telegraph noise may not be a directly implementable noise model in all bosonic systems\footnote{An open quantum system can couple to telegraph noise if it interacts with a fermionic bath; see \cite{abel2008decoherence,franco2012entanglement} for related studies.}, we employ it here as a convenient testing ground. Its well-characterized statistical properties \cite{bena2006dichotomous} and ease of numerical simulation make it a useful drive model for probing how redundant and synergistic information encoding emerges in quantum dynamics. In Section \ref{sec:slow_modes}, we also compare results with those obtained using different noise models to assess their generality.

\subsection{From quantum to semiclassical (mean-field) dynamics}

Directly simulating the Lindblad equation is practical only when the average photon number is small, as the Hilbert space dimension grows rapidly with photon occupation. In this low-photon regime, we simulate full quantum dynamics in \eqref{eqn:Lindblad} to capture all quantum correlations, compute PID and QMI, and analyze how nonclassical effects influence information encoding.

As we increase the driving strength or adjust parameters to reach higher photon-number regimes, the full quantum simulation becomes computationally expensive. In this regime, quantum fluctuations often play a less significant role, and a semiclassical approximation becomes suitable. By factorizing expectation values as \(\langle \hat{a}_i\hat{a}_j \rangle \approx \langle \hat{a}_i\rangle \langle \hat{a}_j\rangle\), the dynamics reduce to coupled nonlinear ordinary differential equations (ODEs) for \(\alpha_i(t)=\langle\hat{a}_i\rangle\):
\begin{align}
\begin{split}
\frac{d}{dt}\alpha_1 &= -(\gamma + i\Delta)\alpha_1 - iJ\alpha_2 - iU_1\alpha_1|\alpha_1|^2 - iF(t), \\
\frac{d}{dt}\alpha_2 &= -(\gamma + i\Delta)\alpha_2 - iJ\alpha_1 - iU_2\alpha_2|\alpha_2|^2 - iF(t).
\end{split}
\label{eqn:meanfieldEOM}
\end{align}
 These ODEs are much easier to solve, allowing us to examine information encoding under conditions where the photon number is large and quantum correlations are negligible.

To bridge the gap between the fully quantum and semiclassical treatments, we also employ a second-order cumulant expansion (see \ref{app:cumulant}). This approach partially restores some quantum correlations while remaining more tractable than the full density-matrix simulation. We expect that in parameter regimes where quantum correlations matter, the cumulant expansion will improve upon the semiclassical approximation, but still remain simpler than the full quantum approach.

\subsection{Characterizing Information Processing}
\label{subsec:info_processing}

To characterize how our system of coupled Kerr-nonlinear oscillators processes and encodes the {\it input} signal $s(t)$ into the {\it output} readouts taken to be
\begin{equation}
X_i(t) \equiv \Re\langle \hat{a}_i(t)\rangle, 
\end{equation}
we analyze three complementary figures of merit. First, we use the partial information decomposition (PID) to separate the total information that the output observables encode about the input into {\it redundant} and {\it synergistic} components. Second, we employ the quantum mutual information (QMI) to quantify the role of quantum correlations in shaping these encoding modes. Finally, we consider the memory capacity in a quantum reservoir computing (QRC) context to assess how information is retained over time. While PID and QMI directly characterize the system’s response to external inputs without any training procedure, the memory capacity inherently involves a training step to quantify how well past inputs can be reconstructed from the system's outputs. In this way, all three measures together provide a comprehensive view of the system’s information processing capabilities.
\newline
\noindent {\it Partial Information Decomposition (PID)}.

Let $s$, $X_1$, and $X_2$ be random variables representing the input signal and the measured observables from the two oscillators, respectively. The mutual information $I(s:X_1,X_2)$ can be decomposed into redundant, synergistic, and unique components as 
\begin{equation}\label{eq: pid_main}
I(s:X_1,X_2) \;=\; \mathrm{Rdn} + \mathrm{Syn} 
\;+\; \mathrm{Unq}(X_1) + \mathrm{Unq}(X_2),
\end{equation}
where $ \mathrm{Rdn}$ is the redundant information present in both $X_1$ and $X_2$, $\mathrm{Unq}(X_i)$ is the unique information contributed solely by $X_i$, and $\mathrm{Syn}$  is the synergistic information accessible only through the joint knowledge of $X_1$ and $X_2$.

By constructing the empirical joint distribution $P(s,X_1,X_2)$ from simulation data and applying the \texttt{BROJA-2PID} algorithm \cite{makkeh2018broja}, we isolate $ \mathrm{Rdn}$ and $\mathrm{Syn}$. This allows us to determine whether the oscillators encode input information redundantly or synergistically, thereby shedding light on their cooperative information processing strategies across different dynamical phases of the system. More detailed discussions and example calculations of PID can be found in \ref{sec:PID} and \ref{app:PID}.

\noindent {\it Quantum Mutual Information (QMI)}.

Although our primary partial information decomposition (PID) analysis focuses on
classical correlations between input-output variables, the underlying reservoir
dynamics remain fundamentally quantum.  To probe quantum correlations inherent in our system,
we compute the \emph{quantum mutual information} (QMI) between the two oscillators.

Let $\rho_{12}$ be the density matrix describing the joint state of the two coupled
Kerr oscillators. We partition the system into subsystems $1$ and $2$, each
corresponding to one oscillator. The QMI is then given by
\begin{equation}
    I(1 : 2) \;=\; S(\rho_1)\; + \; S(\rho_2)\; - \; S(\rho_{12}),
    \label{eq:QMI_def}
\end{equation}
where $\rho_1 = \mathrm{Tr}_2(\rho_{12} )$ and $\rho_2 = \mathrm{Tr}_1(\rho_{12} )$ are the
reduced density matrices of each oscillator, and $S(\rho ) = -\mathrm{Tr}\!
\bigl(\rho  \ln \rho  \bigr)$ is the von Neumann entropy.

Because each oscillator’s Hilbert space is, in principle, infinite-dimensional,
we employ a finite photon-number basis up to a cutoff $N_\mathrm{cutoff}$ to
ensure computational tractability. Namely,
\begin{equation}
    \rho_{12} \;\approx\; \sum_{n_1=0}^{N_\mathrm{cutoff}} \sum_{n_2=0}^{N_\mathrm{cutoff}} 
    c_{n_1,n_2}\, |n_1\rangle\langle n_1|\;\otimes\;|n_2\rangle\langle n_2|,
    \label{eq:rho_fock}
\end{equation}
where $|n_i\rangle$ is the Fock state with $n_i$ photons in oscillator $i$,
and $c_{n_1,n_2}$ are the matrix elements obtained by time-averaging
the steady-state solution of the master equation. We verify that increasing
$N_\mathrm{cutoff}$ beyond $\sim 10$ does not significantly alter the results 
within the quantum regime studied in this work, suggesting that the computed QMI 
converges to within numerical precision.

To evaluate Eq.~(\ref{eq:QMI_def}), we first diagonalize the full two-oscillator 
density matrix $\rho_{12}$
\begin{equation}
    \rho_{12} \;=\; \sum_{k} \lambda_k \,\bigl|\psi_k\bigr\rangle 
    \bigl\langle \psi_k\bigr|,
    \quad
    S(\rho_{12}) \;=\; -\sum_{k} \lambda_k \,\ln \lambda_k,
\end{equation}
where $\{\lambda_k\}$ and $\{\lvert \psi_k\rangle\}$ are the eigenvalues and 
eigenbases of $\rho_{12}$, respectively. We then obtain the reduced density 
matrices for each oscillator by tracing out the other as
$\rho_1 = \mathrm{Tr}_2 \bigl(\rho_{12}\bigr),$ and  $\rho_2 = \mathrm{Tr}_1 \bigl(\rho_{12}\bigr).$
Both $\rho_1$ and $\rho_2$ are similarly diagonalized to compute their von Neumann 
entropies, $S(\rho_1)$ and $S(\rho_2)$. Substituting these into 
Eq.~(\ref{eq:QMI_def}) gives the QMI. The resulting $I(1 : 2)$ quantifies the \emph{total} correlations between the two
oscillators, including both classical and quantum components, and thus 
complements a classical PID-based analysis.

\section{Results and Discussion}
\label{sec:results}

We now present numerical evidence that coupled Kerr oscillators can encode input signals in either a redundant or synergistic fashion, depending on $J$, $\Delta$, and $\gamma$. 

\subsection{Emergence of Synergistic Encoding}
\label{sec:emergence_synergy}

We begin by examining how the joint mutual information (MI), $I\bigl(s:(X_1,X_2)\bigr)$, compares to the individual MIs $I(s:X_1)$ and $I(s:X_2)$ when probing our driven-dissipative system. We focus on two representative parameter sets: a \emph{mean-field} regime with larger drive and smaller Kerr nonlinearities, and a \emph{quantum} regime with smaller drive and larger Kerr nonlinearities. In both cases, we fix the detuning and damping at $\Delta = -2$ and $\gamma = 0.5$. Concretely, in the mean-field case, we take $F=2.0$, $U_1=6.25\times10^{-3}$, and $U_2=2U_1$, while in the quantum regime we take $F=0.2$, $U_1=4.0$, and $U_2=2U_1$. 

\begin{figure}[!ht]
\begin{center}
\includegraphics[keepaspectratio, width=0.47\textwidth]{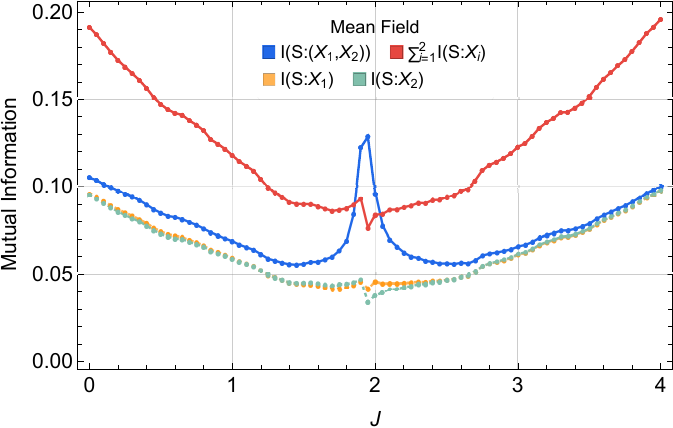}
\includegraphics[keepaspectratio, width=0.47\textwidth]{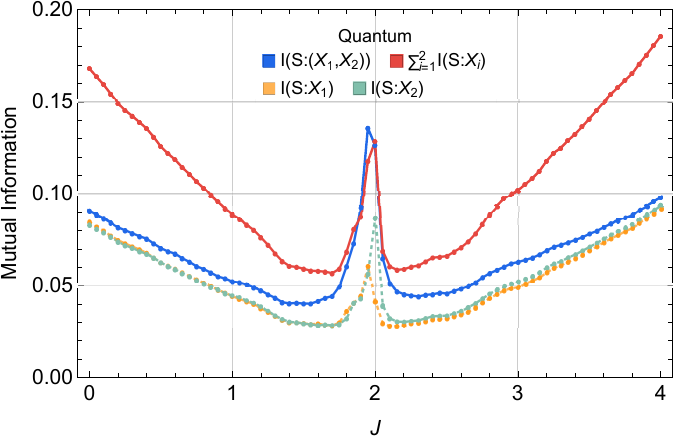}
\caption{Classical mutual information $I\bigl(s:(X_1,X_2)\bigr)$, compared to $I(s:X_1)$ and $I(s:X_2)$ in (left) a mean-field regime and (right) a quantum regime. In the mean field regime, $I\bigl(s:(X_1,X_2)\bigr)$ exceeds $I(s:X_1)$ or $I(s:X_2)$ alone near $J = |\Delta|$, hinting at synergy. On the other hand, in the quantum regime, $I\bigl(s:(X_1,X_2)\bigr)$ is comparable to, but not always exceeding, the sum of  $I(s:X_1)$ and $I(s:X_2)$.}
\label{fig:prototype_result}
\end{center}
\end{figure}

\paragraph{Synergy from total mutual information consideration.}
Figure~\ref{fig:prototype_result} compares \newline $I\bigl(s:(X_1,X_2)\bigr)$ with $I(s:X_1)$ and $I(s:X_2)$, revealing that
\[
I\bigl(s:(X_1,X_2)\bigr) \;>\; I(s:X_1)\; + \; I(s:X_2),
\]
in the mean-field dynamics regime. This information excess indicates that measuring both $X_1$ and $X_2$ jointly can reveal strictly more information about the external drive signal $s(t)$ than either observable alone, suggesting a potential \emph{synergistic} encoding mechanism.

\paragraph{Transition from redundant to synergistic encoding.}
To further investigate whether this information surplus really originates from synergistic effects (rather than unique information in each oscillator), we perform partial information decomposition (PID) according to Eq.~\eqref{eq: pid_main}.
For clarity, we normalize synergy and redundancy by $I\bigl(s:(X_1,X_2)\bigr)$, respectively, 
\[
\mathrm{Syn}_{\mathrm{norm}} = \frac{\mathrm{Syn}}{I\bigl(s:(X_1,X_2)\bigr)},
\quad
\mathrm{Rdn}_{\mathrm{norm}} = \frac{\mathrm{Rdn}}{I\bigl(s:(X_1,X_2)\bigr)}.
\]

\begin{figure}[!ht]
\begin{center}
\begin{tabular}{cc}
\begin{subfigure}{0.47\textwidth}
\includegraphics[width=\textwidth]{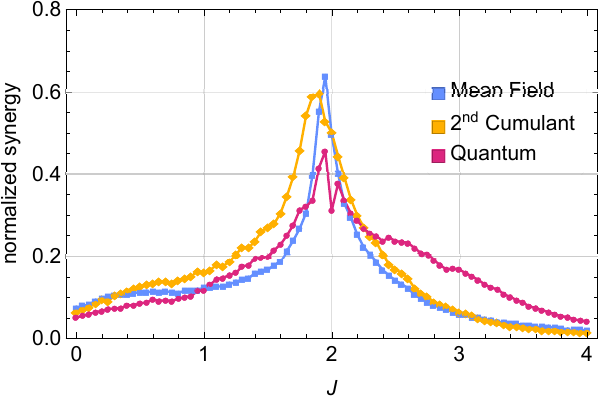}
\end{subfigure} &
\begin{subfigure}{0.47\textwidth}
\includegraphics[width=\textwidth]{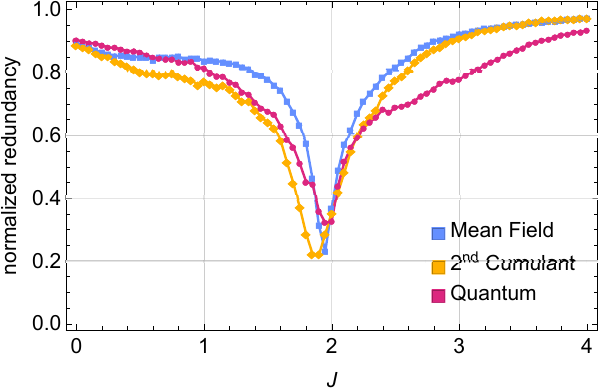}
\end{subfigure}
\end{tabular}
\caption{Normalized synergy (left) and normalized redundancy (right) vs.\ the coupling $J$. A pronounced peak near $J \approx |\Delta|$ marks the crossover from predominantly redundant to more synergistic encoding. In the fully quantum description, enhanced quantum correlations can favor redundancy even at the transition, whereas second-order cumulants interpolate between mean-field and quantum descriptions.}
\label{fig:compare05}
\end{center}
\end{figure}

In Fig.~\ref{fig:compare05}, we compare the normalized synergy (left) and redundancy (right) across three regimes (mean-field, second-order cumulant, and fully quantum) at fixed detuning and dissipation. As we sweep the coupling strength $J$ from small to large, a pronounced synergy peak emerges near $J \approx |\Delta| = 2$, marking a transition from predominantly redundant encoding to notably higher synergy (near $J \simeq |\Delta|$). In the quantum regime, stronger quantum correlations bias the encoding scheme slightly toward redundancy even near the peak. In contrast, the second-order cumulant approach captures the partial quantum correlations that lie between the mean-field regime (where correlations are suppressed) and the fully quantum regime (where all orders of correlations may appear). This second-order cumulant dynamics provides an approximate interpolation for regimes where quantum correlations are significant but not too strong\footnote{We set $\Delta = -2, F = 0.5, \gamma = 0.5, U_1 = 0.2, U_2 = 2U_1$ to represent a dynamical regime with non-negligible quantum correlations, motivating the use of second-order cumulants}, see~\ref{app:cumulant}.

\subsection{Underlying Mechanisms Enhancing Synergistic Coding: The Role of Soft and Fast Modes}
\label{sec:slow_modes}

In this section, we explain the sharp increase in synergy observed near $J \simeq |\Delta|$ and attribute this behavior to the interplay between soft and fast modes in the coupled Kerr oscillators. Specifically, we demonstrate that the dominance of fast modes, enabled by the overdamping of soft modes, enhances coherent collective dynamics, leading to an increase in synergistic information.

\paragraph{Soft modes and potential landscape flatness.}

\begin{figure}[!ht]
\begin{center}
\includegraphics[width = 0.95\textwidth]{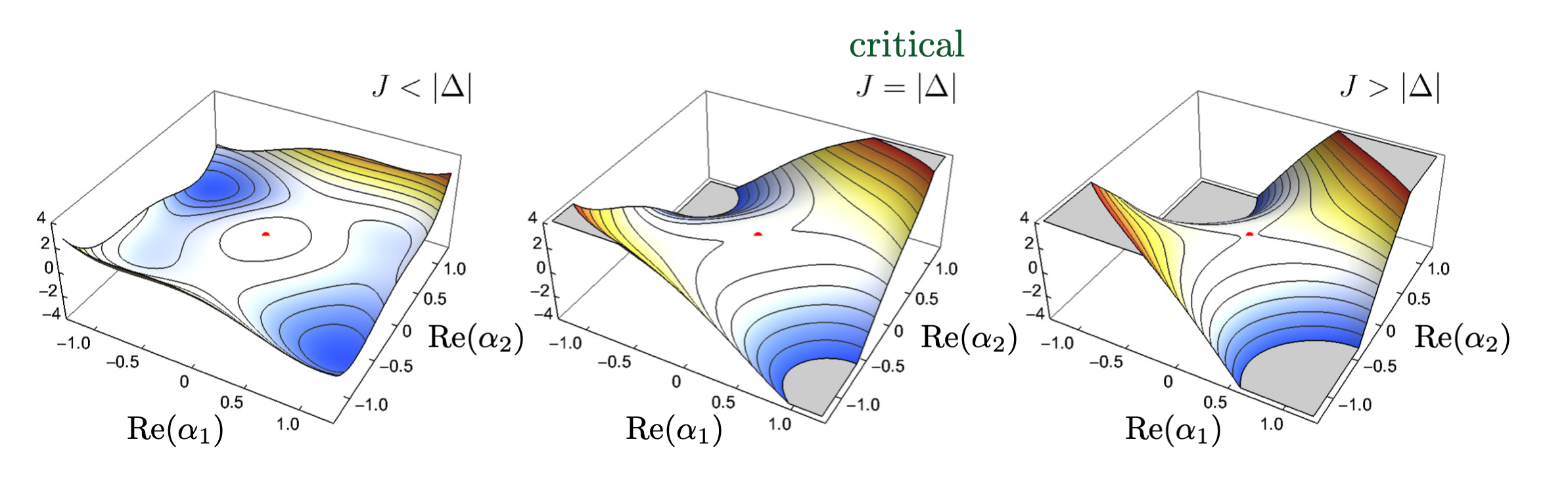}
\caption{Effective potential around the steady state $\alpha_{1,c}=\alpha_{2,c}=0$ (red dot) in the $\Delta < 0, J > 0$ regime, projected onto $\mathrm{Im}(\alpha_1)=\mathrm{Im}(\alpha_2)=0$. \textbf{(Left)} When $J < |\Delta|$, the steady state is weakly unstable in all directions, 
with no soft modes present, and the system predominantly encodes inputs redundantly. 
\textbf{(Center)} At the critical point $J = |\Delta|$, flat directions appear, marking 
the onset of soft modes. In this near-critical regime, collective oscillations enhance 
synergistic encoding. 
\textbf{(Right)}  For $J > |\Delta|$, the potential deforms into a saddle, with two stable 
and two unstable directions. Here, the soft modes again disappear, and the system 
encodes inputs redundantly. }
\label{fig:potential_landscape}
\end{center}
\end{figure}

In the mean-field approximation without external drive ($F=0$), the coupled Kerr oscillators evolve in the following effective potential (see \ref{app:Keldysh}),  which captures the interplay of detuning, Kerr nonlinearities, and coupling
\begin{equation}\label{eqn:Veff} 
V\bigl(\vec{\alpha}_c,\vec{\alpha}_c^*\bigr)
\,=\,
\sum_{j=1,2}
\Bigl(\Delta\,|\alpha_{j,c}|^2 + \tfrac{1}{4}U_j\,|\alpha_{j,c}|^4\Bigr)
\;+\;
J\bigl(\alpha_{1,c}\,\alpha_{2,c}^* + \alpha_{2,c}\,\alpha_{1,c}^*\bigr),
\end{equation}
where $\Delta_1 = \Delta_2 = \Delta$ and $\vec{\alpha}_c = [\alpha_{1,c}, \alpha_{2,c}]$. At $J = |\Delta|$, the Hessian of $V$ evaluated at the steady-state solution $\alpha_{1,c} = \alpha_{2,c} = 0$ develops zero eigenvalues, corresponding to flat or marginal directions. These flat directions represent soft modes, characterized by near-zero oscillation frequencies. Specifically, when evaluated at the steady state $\alpha_{1,c} = \alpha_{2,c} = 0$, the Hessian matrix of this effective potential \eqref{eqn:Veff} calculated in terms of the vector $ (\Re(\alpha_{1,c}), \Im(\alpha_{1,c}), \Re(\alpha_{2,c}), \Im(\alpha_{2,c}))$ is
\begin{equation}
    \mathbf{H}[V(\vec{\alpha}, \vec{\alpha}^*)]|_{\vec{\alpha}, \vec{\alpha}^* = 0} = 2\mqty( \Delta & 0 & J & 0 \\
    0 & \Delta & 0 & J \\
    J & 0 & \Delta & 0 \\
    0 & J & 0 & \Delta).
\end{equation}
The eigenvalues of this Hessian are
\begin{equation*}
\{\: 2(\Delta+J),\: 2(\Delta+J),\: 2(\Delta-J),\: 2(\Delta-J)\:\}. 
\end{equation*}
For the relevant parameter regime in which $\Delta < 0$ and $J > 0$, exactly when $J = |\Delta|$ that two flat directions (zero eigenvalues) emerge, and the other two directions are unstable at the second order (negative eigenvalues), see Fig.~\ref{fig:potential_landscape}.

\paragraph{Overdamping of soft modes at $J = |\Delta|$.}

Including dissipation with a rate $\gamma$ can transform the nearly flat directions into overdamped dynamics as follows.  Following the Keldysh formalism \cite{soriente2020distinguishing,soriente2021distinctive,alaeian2021noise} (see \ref{app:Keldysh}), the retarded Green’s function shows poles of the form
\begin{equation}\label{eq:G_retarded_poles}
\omega_s \;=\; \pm\!\bigl|J - |\Delta|\bigr|\;-\;i\,\gamma,
\quad
\omega_f \;=\; \pm\!\bigl|J + |\Delta|\bigr|\;-\;i\,\gamma,
\end{equation}
where $\omega_s$ (slow) and $\omega_f$ (fast) label the respective branches. Exactly at $J=|\Delta|$, the real part of $\omega_s$ \emph{vanishes}, leaving only $-\,i\,\gamma$, indicating an overdamped relaxation to the steady state. In contrast, the fast modes remain oscillatory with frequencies $\mathrm{Re}(\omega_f) = |J + |\Delta||$. Consequently, at $J = |\Delta|$, the dynamics are dominated by the coherent oscillations of the fast modes, as the soft modes contribute only non-oscillatory relaxation.

\paragraph{Coherence-driven synergy enhancement.}

When the dissipation rate is comparable to the oscillatory frequency of the fast modes ($\gamma \sim \Re (\omega_f)$), the dominance of fast modes at $J = |\Delta|$ reduces competition between oscillatory frequencies\footnote{Soft modes disappear and only fast modes persist.} and enhances the coherence of the system's collective response. More specifically, following the discussion in \ref{app:Keldysh}, one can consider the relaxation dynamics of the small perturbation $\delta \alpha_c(t)$ around the steady state. It follows that the relaxation dynamics of each observable at site $j$ can be expressed as
\begin{equation}
\Re \bigl( \delta \alpha_{j,c}(t) \bigr) = c_{j,s} e^{-\gamma t} \cos(\Re (\omega_s) t) + c_{j,f} e^{-\gamma t} \cos(\Re(\omega_f) t),
\end{equation}
where $c_{j,s}$ and $c_{j,f}$ are initial-perturbation-dependent mode amplitudes. When $\mathrm{Re}(\omega_s) \to 0$, the soft mode contribution simplifies to pure exponential decay, and the dissipative dynamics are dominated by the single oscillatory frequency $\omega_f$ of the fast modes.

This coherence relaxation eliminates competing oscillations and minimizes overlapping contributions between modes, reducing redundancy. Also, collective dynamics driven by the fast modes encode information that cannot be captured by any single oscillator alone, thereby enhancing synergistic information. This explains the peak in synergy observed near $J \simeq |\Delta|$.

\begin{figure}[!ht]
\begin{center}
\includegraphics[width = \textwidth]{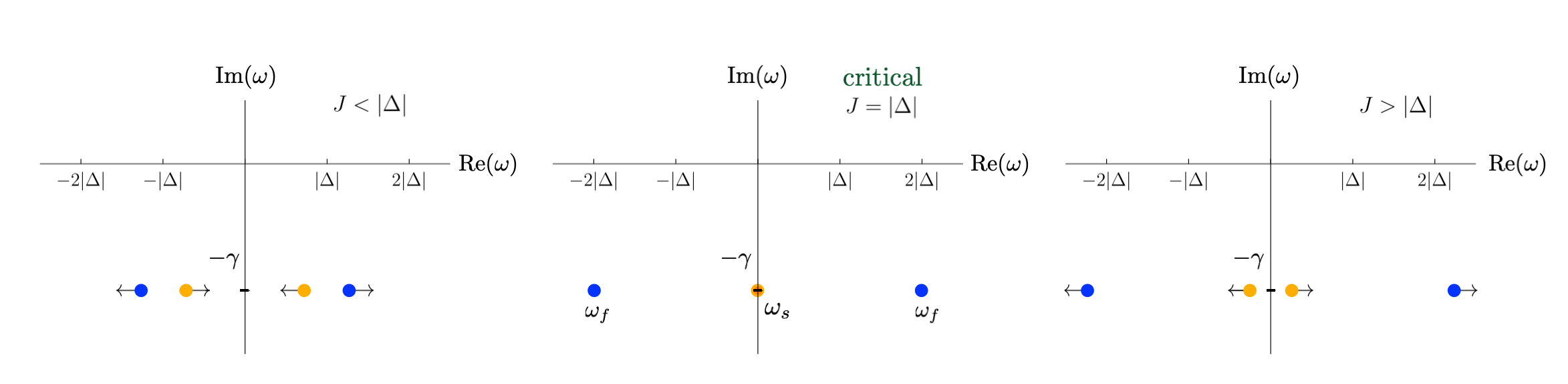}
\caption{Retarded Green’s function poles~\eqref{eqn:Greenfunction} in the complex-frequency plane as $J$ increases, 
illustrating the evolution of slow modes $\omega_s$ (orange dots) and fast modes $\omega_f$ (blue dots). 
For $J < |\Delta|$, both slow and fast modes coexist, and the system tends to encode inputs 
more redundantly. Near the critical point $J = |\Delta|$, the real part of $\omega_s$ approaches 
zero, indicating a disappearance of slow collective oscillations and an overdamped decay. 
In this near-critical regime, collective oscillations due to underdamped fast modes dominate and enhance synergistic 
encoding. For $J > |\Delta|$, the slow modes shift away from zero frequency, reducing synergy 
and transitioning the system back toward redundant encoding. This interplay between 
slow and fast modes governs how the system transitions from predominantly redundant to 
synergistic processing and back again as $J$ crosses the critical point.}
\label{fig:omega_varyJ}
\end{center}
\end{figure}

Figure~\ref{fig:omega_varyJ} illustrates how the slow-mode poles (orange) move to the imaginary axis at $J=|\Delta|$, marking the \emph{disappearance of competing oscillatory timescales}. In this near-critical regime, the relaxation dynamics become dominated by the underdamped (oscillatory) contributions of the fast modes, resulting in coherent dissipation. It is important to note that this result pertains to the regime where $\gamma \sim |\Re(\omega_f)| \gg |\Re(\Omega_s)|$. 

\paragraph{Generality of the transition to synergistic behavior.}

\begin{figure}[!ht]
\begin{center}
\begin{tabular}{cc}
\begin{subfigure}{0.47\textwidth}
\includegraphics[keepaspectratio, width = \textwidth]{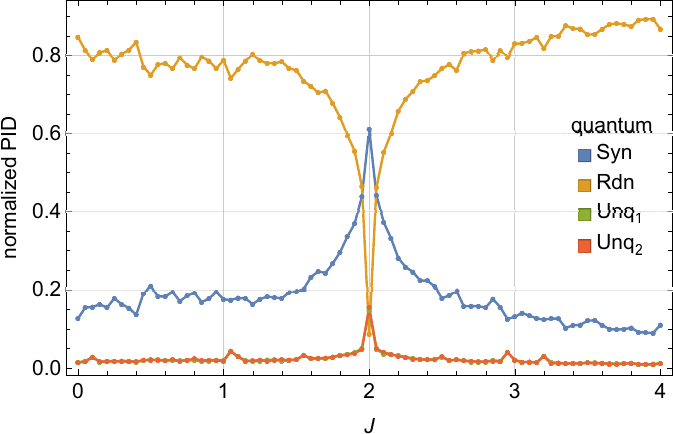}
\caption{Normalized PID with uniform noise input.}
\end{subfigure} &
\begin{subfigure}{0.47\textwidth}
\includegraphics[keepaspectratio, width = \textwidth]{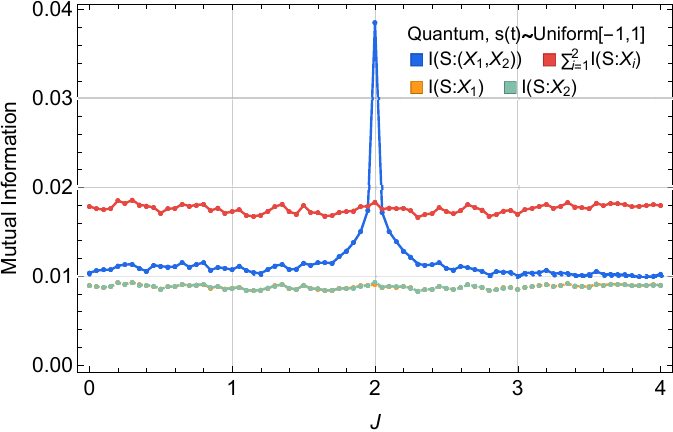}
\caption{Comparison of total MI and partial MI contributions.}
\end{subfigure}
\end{tabular}
\caption{Impact of uniform, uncorrelated noise input on information encoding at the quantum regime (\(\Delta = -2\), \(\gamma = 0.5\), and \(F = 0.2\)). (a) Normalized PID components averaged over 50 noise realizations show a clear transition to synergistic encoding near \(J = |\Delta|\). (b) Comparison between the total mutual information \(I(S:(X_1,X_2))\) and the sum of individual mutual information contributions \(I(S:X_1) + I(S:X_2)\), highlighting the emergence of synergy near the critical coupling.}
\label{fig:Info_uniform}
\end{center}
\end{figure}

The observed synergy peak at \(J \simeq |\Delta|\) is not specific to the type of input signal driving the system. To demonstrate this, we performed numerical simulations of the master equation describing quantum dynamics with the input signal \(s(t)\) sampled from a uniform distribution in the interval \([-1,1]\), and uncorrelated in time. The results, shown in Fig.~\ref{fig:Info_uniform}, confirm that the transition in redundant/synergistic behavior persists at \(J = |\Delta|\), regardless of the statistical properties of the input.

These results emphasize that the transition to synergistic encoding at \(J \simeq |\Delta|\) is an intrinsic feature of the system’s response dynamics, driven by the dominance of underdamped fast modes, and not much by the input signal properties. In the following section, we explore how increasing \(\gamma\) impacts the system’s encoding behavior, showing that fast relaxation towards the steady state progressively shifts the system from synergistic to redundant encoding.



\subsection{Large Dissipation Leads to Redundant Encoding} \label{sec:role_gamma}

As the damping rate \(\gamma\) increases, the dynamics progressively shift toward overdamped relaxation for both slow and fast modes. This transition leads to a steady-state regime where the two cavities become nearly identical, resulting in redundant coding of the input information. While the emergence of soft modes and the dominance of fast modes at \(J \simeq |\Delta|\) enhance synergy, increasing \(\gamma\) gradually suppresses this effect, shifting the system toward a more redundant encoding regime dominated by rapid overdamped relaxation towards the steady state.

\begin{figure}[!ht]
\begin{center}
\begin{tabular}{cc}
\begin{subfigure}{0.45\textwidth}
\includegraphics[keepaspectratio, width = \textwidth]{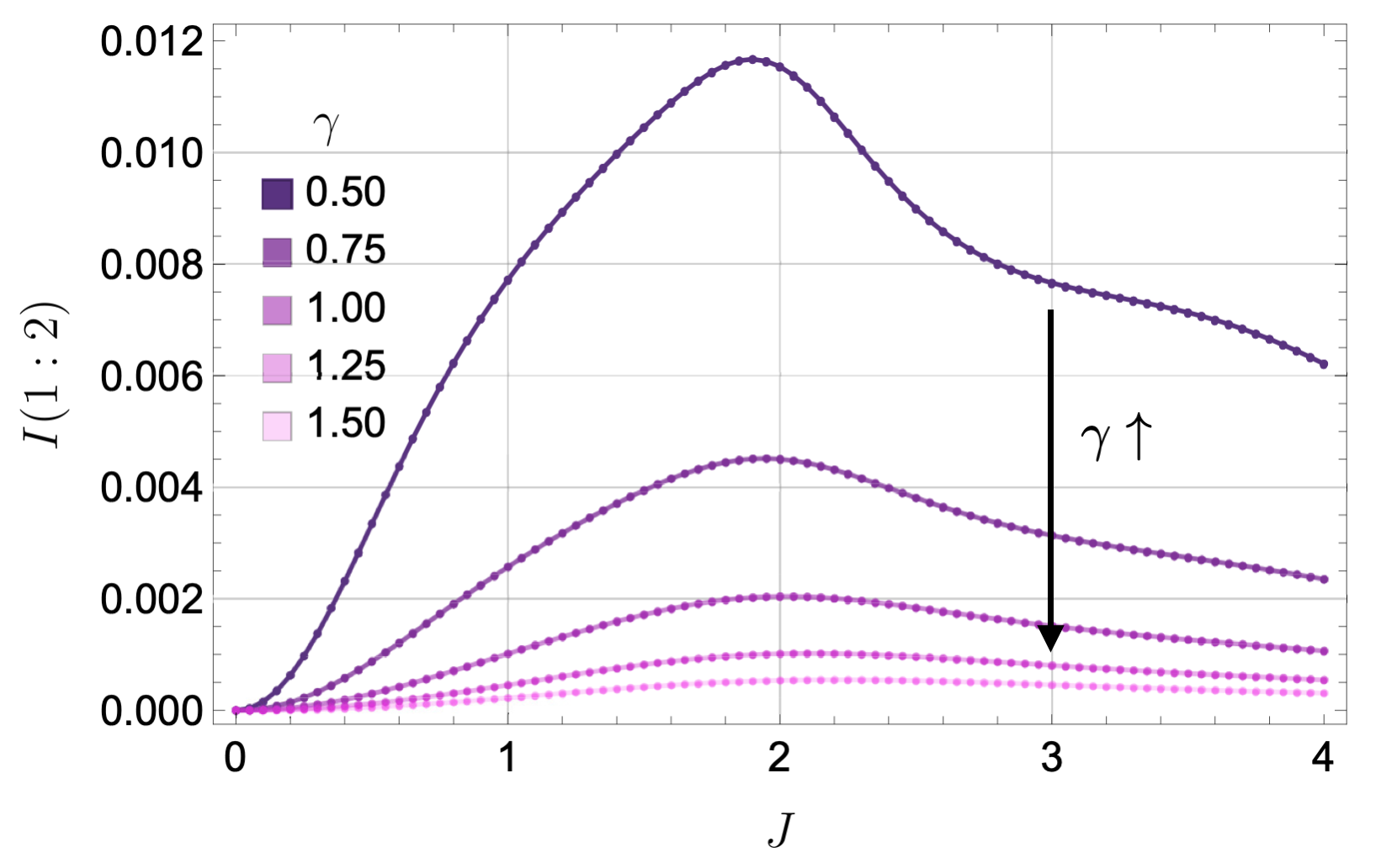}
\caption{QMI vs $J$ for different $\gamma$.}
\end{subfigure} &
\begin{subfigure}{0.45\textwidth}
\includegraphics[keepaspectratio, width = \textwidth]{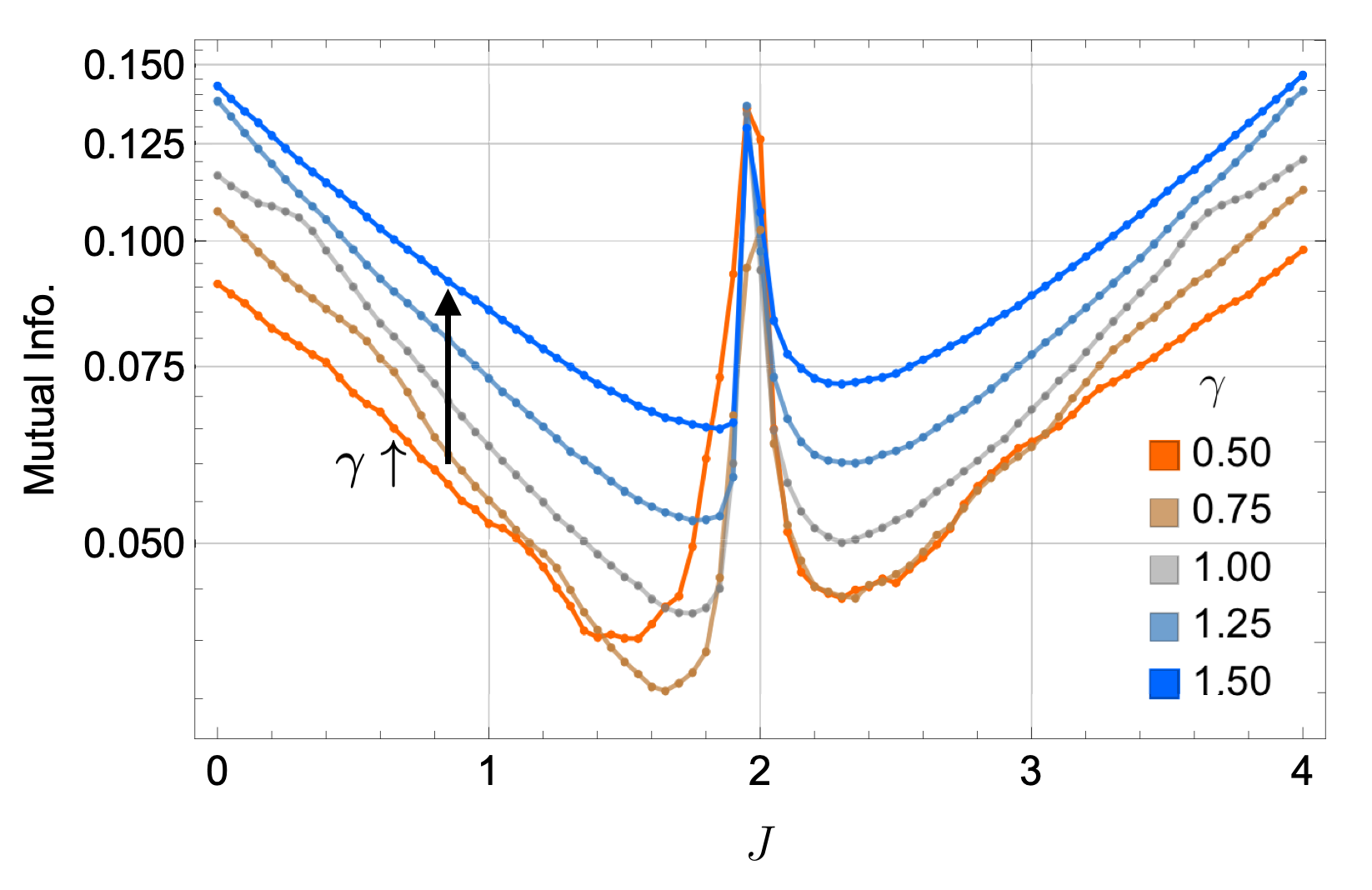}
\caption{Classical MI vs $J$ for different $\gamma$.}
\end{subfigure} \\
\begin{subfigure}{0.45\textwidth}
\includegraphics[keepaspectratio, width = \textwidth]{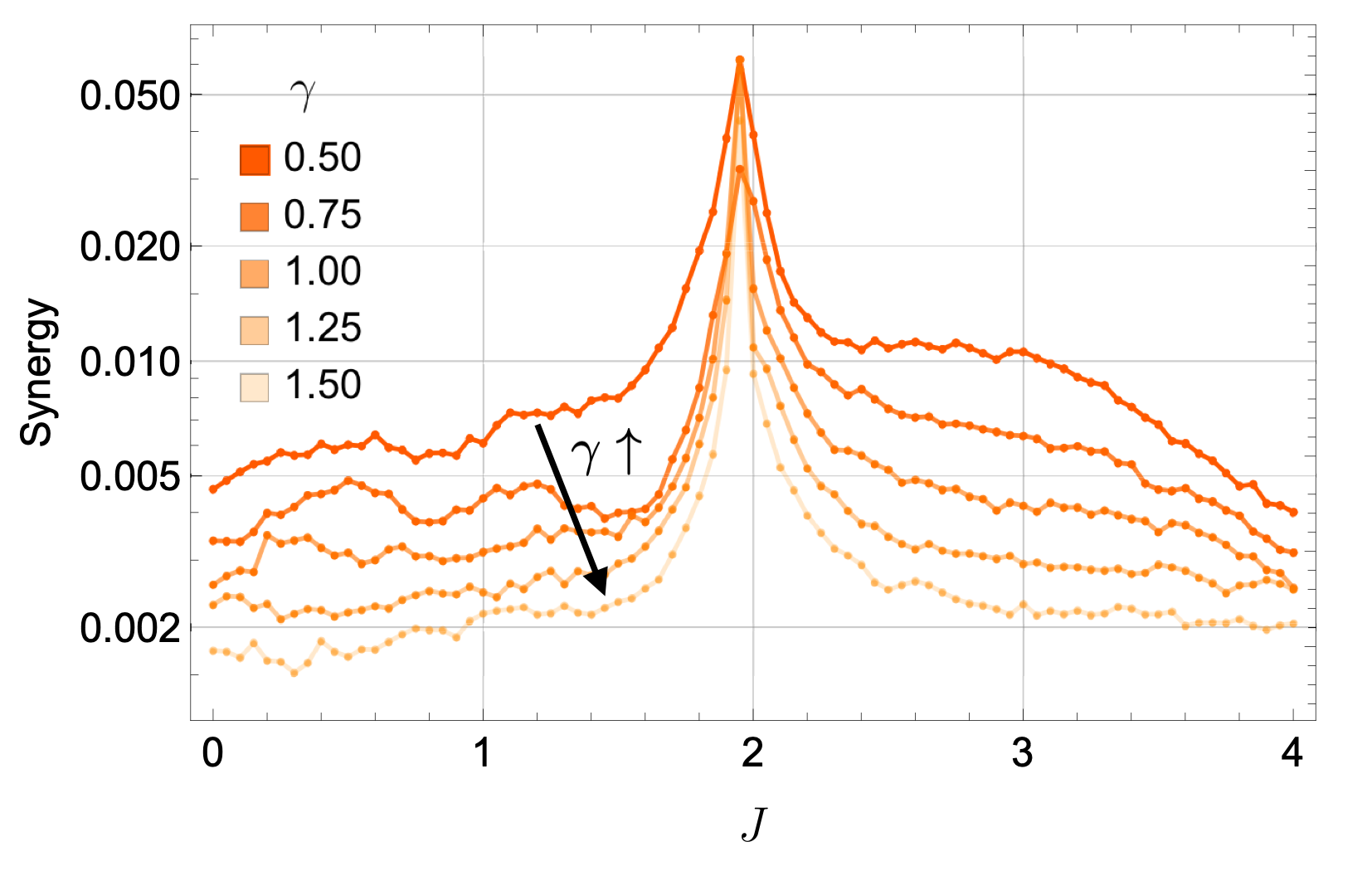}
\caption{Synergy vs $J$ for different $\gamma$.}
\end{subfigure} &
\begin{subfigure}{0.45\textwidth}
\includegraphics[keepaspectratio, width = \textwidth]{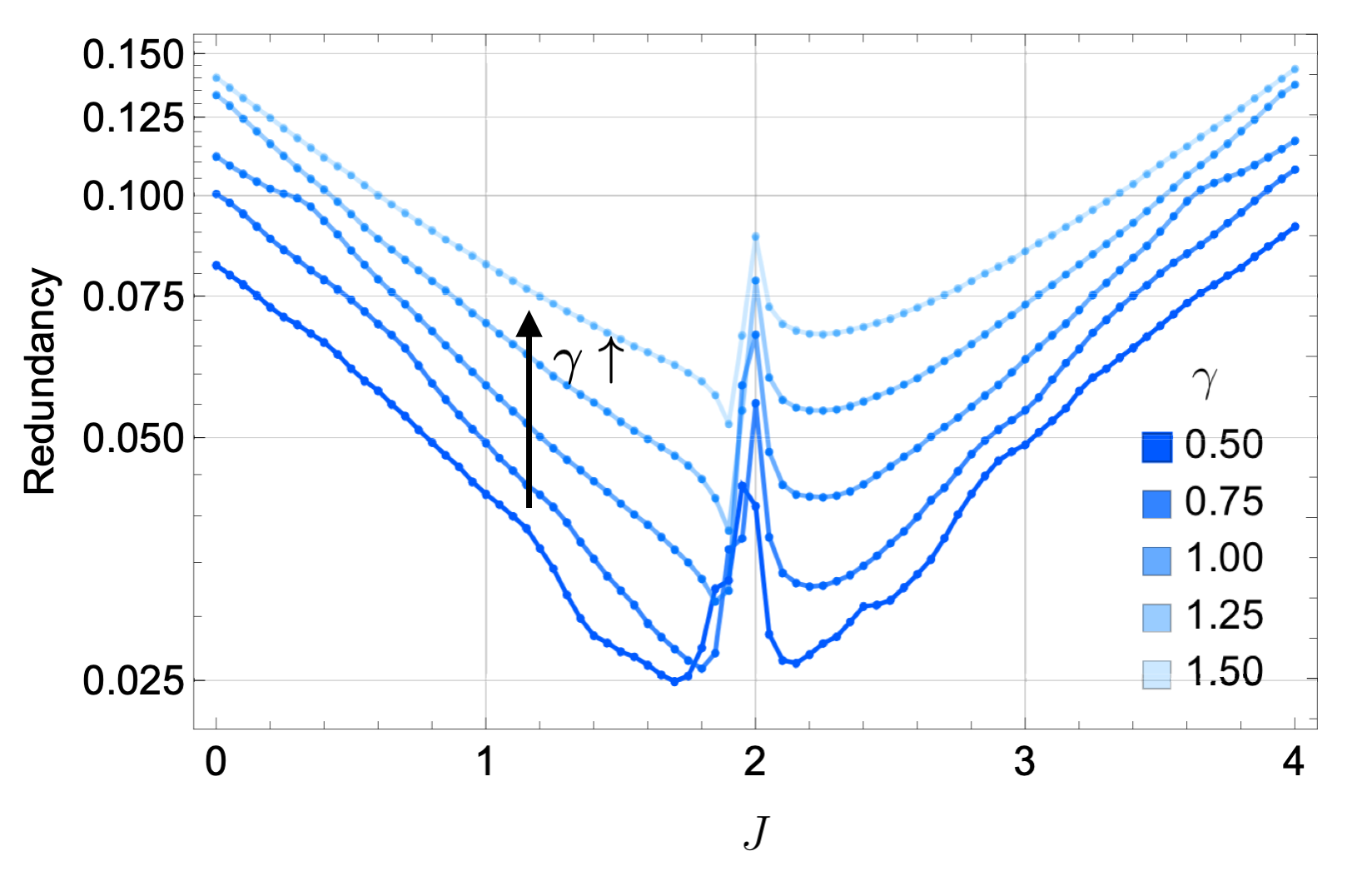}
\caption{Redundancy vs $J$ for different $\gamma$.}
\end{subfigure}
\end{tabular}
\caption{Impact of increasing \(\gamma\) on information metrics in the quantum dynamics regime (\(\Delta = -2\) and \(F = 0.2\)). 
(a) Time-averaged quantum mutual information (QMI) between the two oscillators as a function of \(J\). 
(b) Classical mutual information (MI) between the input signal and the output observables as a function of \(J\). 
(c) Absolute synergy, and (d) absolute redundancy, both plotted as functions of \(J\). 
These plots illustrate the transition from low-synergistic to high-redundant encoding with increasing \(\gamma\). 
At large dissipation, the two subsystems approach a product state, becoming effectively independent, as indicated by the low QMI. 
Interestingly, despite redundancy dominating at higher \(\gamma\), the total mutual information at \(J = |\Delta|\) near criticality remains approximately constant.}
\label{fig:combined_gamma_effect}
\end{center}
\end{figure}

This behavior is particularly evident in the quantum regime, where the system approaches a product state at large \(\gamma\), rendering the two oscillators effectively independent. As shown in Fig.~\ref{fig:combined_gamma_effect}, increasing \(\gamma\) from the baseline value of \(\gamma = 0.5\) (as seen in Figs.~\ref{fig:prototype_result} and \ref{fig:compare05}) results in a clear reduction in quantum correlations. Panel (a) of Fig.~\ref{fig:combined_gamma_effect} quantifies this through the time-averaged quantum mutual information (QMI), which decreases as \(\gamma\) increases. Notably, peaks in QMI at \(J \simeq |\Delta|\) coincide with peaks in classical mutual information (MI), as shown in panel (b). This observation aligns with the intuition that higher quantum correlations often translate to enhanced classical information encoding near the critical coupling.
Panels (c) and (d) further illustrate this transition by showing how absolute synergy diminishes and absolute redundancy grows with increasing \(\gamma\). At low \(\gamma\), the dynamics favor (weakly) synergistic encoding. At high \(\gamma\), however, the system becomes dominated by (highly) redundant encoding, with both cavities responding similarly and independently of each other.

\subsection{Memory Capacity of Synergistic and Redundant Encoding}\label{sec:MC}

We close our discussion by analyzing the performance of the coupled Kerr oscillators as a quantum reservoir, focusing on their capacity to retain and process temporal information. This memory capacity benchmark highlights how the synergistic and redundant behaviors identified earlier influence practical tasks such as time-series memorization.

\paragraph{Short-term memory task.} To quantify memory capacity, we train the system to recall past input signals using a short-term memory task. The input signal \(s(t)\) is sampled from a uniform distribution in the interval \([-1, 1]\) and is uncorrelated in time, and the target time series \(\bar{y}_n(t)\) corresponds to the input signal at a previous time step:
\begin{equation}
    \bar{y}_n(t) = s(t - n\Delta t), \label{eqn:memory_task}
\end{equation}
where \(\Delta t = 0.01\) is the time step. The output observables \(X_i(t) = \Re(\langle \hat{a}_i(t) \rangle)\) and \(Y_i(t) = \Im(\langle \hat{a}_i(t) \rangle)\) are used as feature vectors.\footnote[1]{Here we use more output readouts than in the previous sections since this input signal is more difficult to fit with less feature vectors.} We construct an output vector \(\vec{X}(t)\):
\begin{equation}
    \vec{X}(t) = \mqty(X_1(t), X_2(t), Y_1(t), Y_2(t), 1)^\top,
\end{equation}
and fit the target \(\bar{y}_n(t)\) using a weight matrix \(\mathbf{W}\) via standard linear regression with Tikhonov regularization:
\begin{equation}
    \hat{y}(t) = \mathbf{W}^* \vec{X}(t),
\end{equation}
where the optimal weights \(\mathbf{W}^*\) minimize the mean-square error (MSE) during training:
\begin{equation}
    \text{MSE}(\mathbf{W}, \lambda) = ||\bar{\mathbf{y}} - \hat{\mathbf{y}}||^2 + \lambda||\mathbf{W}||^2.
\end{equation}
For simplicity, we set \(\lambda = 0\). After training, the memory capacity  \cite{jaeger2001short} for delay step \(n\) is evaluated as:
\begin{equation}\label{eq:MC}
    \text{MC}(n) = \frac{\text{cov}^2(\bar{\mathbf{y}}_n, \hat{\mathbf{y}}_{\text{pred}})}{\sigma^2(\bar{\mathbf{y}}_n)\sigma^2(\hat{\mathbf{y}}_{\text{pred}})},
\end{equation}
where \(0 \leq \text{MC}(n) \leq 1\), with \(\text{MC}(n) = 1\) indicating perfect recall.

\begin{figure}[!ht]
\begin{center}
\begin{tabular}{ccc}
\begin{subfigure}{0.45\textwidth}
\includegraphics[keepaspectratio, width = \textwidth]{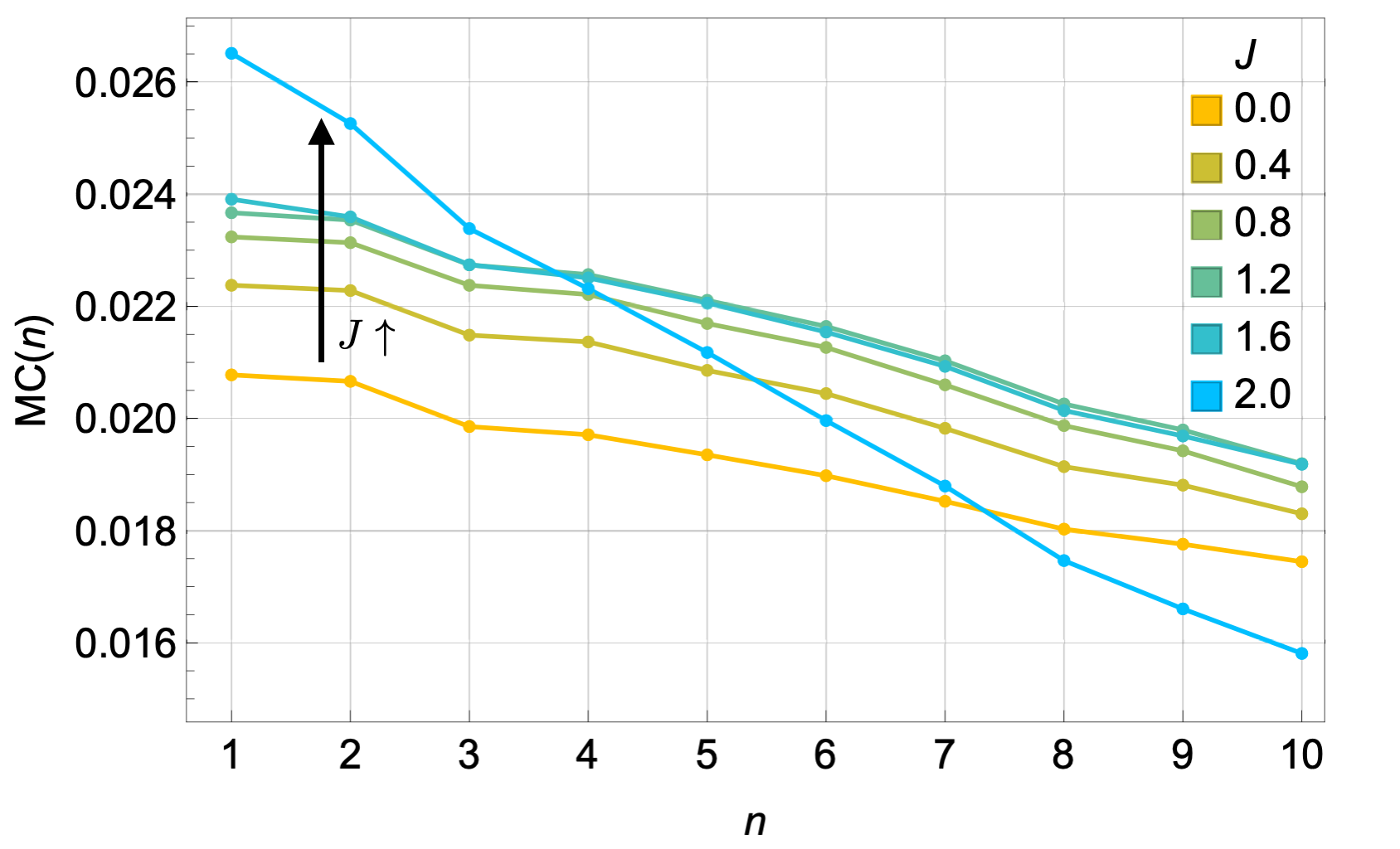}
\caption{\(J \in [0, 2]\)}
\end{subfigure} &
\begin{subfigure}{0.45\textwidth}
\includegraphics[keepaspectratio, width = \textwidth]{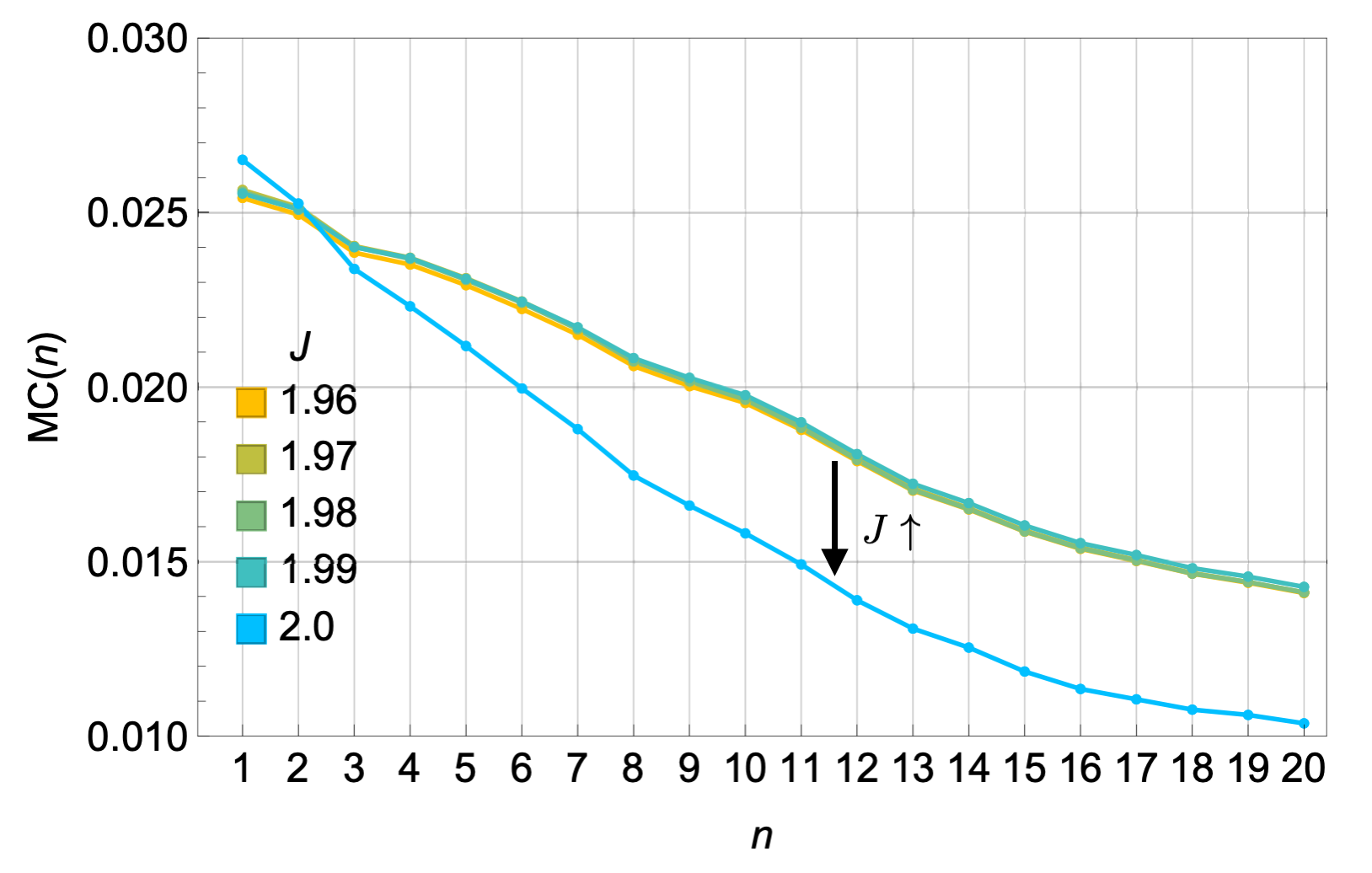}
\caption{\(J \in [1.96, 2]\)}
\end{subfigure}
\end{tabular}
\caption{Memory capacity as a function of \(n\) as \(J\) approaches the critical coupling at $J=|\Delta|$ for the parameters \(\Delta = -2\), \(\gamma = 0.5\), and \(F = 0.2\). (a) \(n = 1{-}10\) for \(J \in [0, 2]\), showing an increase in short-term memory capacity as \(J \rightarrow |\Delta| \). (b)  \(n = 1{-}20\) for \(J \in [1.96, 2]\), showing the long-term capacity drops as \(J \rightarrow |\Delta| \). These results are averaged over 50 input realizations.}
\label{fig:MCvaryJ}
\end{center}
\end{figure}

\paragraph{Tuning \(J\) towards the critical coupling strength.} Figure~\ref{fig:MCvaryJ} shows the memory capacity as \(J\) approaches the critical coupling \(J = |\Delta| = 2\). At smaller values of \(J\), memory capacity remains low across delay steps. However, as \(J\) approaches \(|\Delta|\), the system exhibits a notable change in behavior: memory capacity for short delays (small $n$) increases significantly, while memory capacity for longer delays (large $n$) decays more rapidly. This behavior reflects the trade-off between short-term and long-term memory as the system transitions to the synergistic regime dominated by fast modes. 

We note that, for low dissipation (e.g., \(\gamma = 0.5\)), the system may appear to retain information about initial conditions at long delays \(n\), indicating that the fading memory property might not be fully realized. While a more comprehensive benchmark such as the information processing capacity (IPC) \cite{Dambre2012} could more definitively confirm or refute fading memory in this low-dissipation regime, our current goal is to investigate synergy and redundancy in a small quantum reservoir rather than to perform an exhaustive reservoir-computing benchmark. Instead, we emphasize qualitative trends in short- vs.\ long-term memory. A more detailed, IPC-based study would be an interesting direction for future work to fully study the computational potential of near-critical quantum reservoirs in this system.

\paragraph{Impact of dissipation \(\gamma\).} 

\begin{figure}[!ht]
\begin{center}
\begin{tabular}{cc}
\begin{subfigure}{0.45\textwidth}
\includegraphics[keepaspectratio, width = \textwidth]{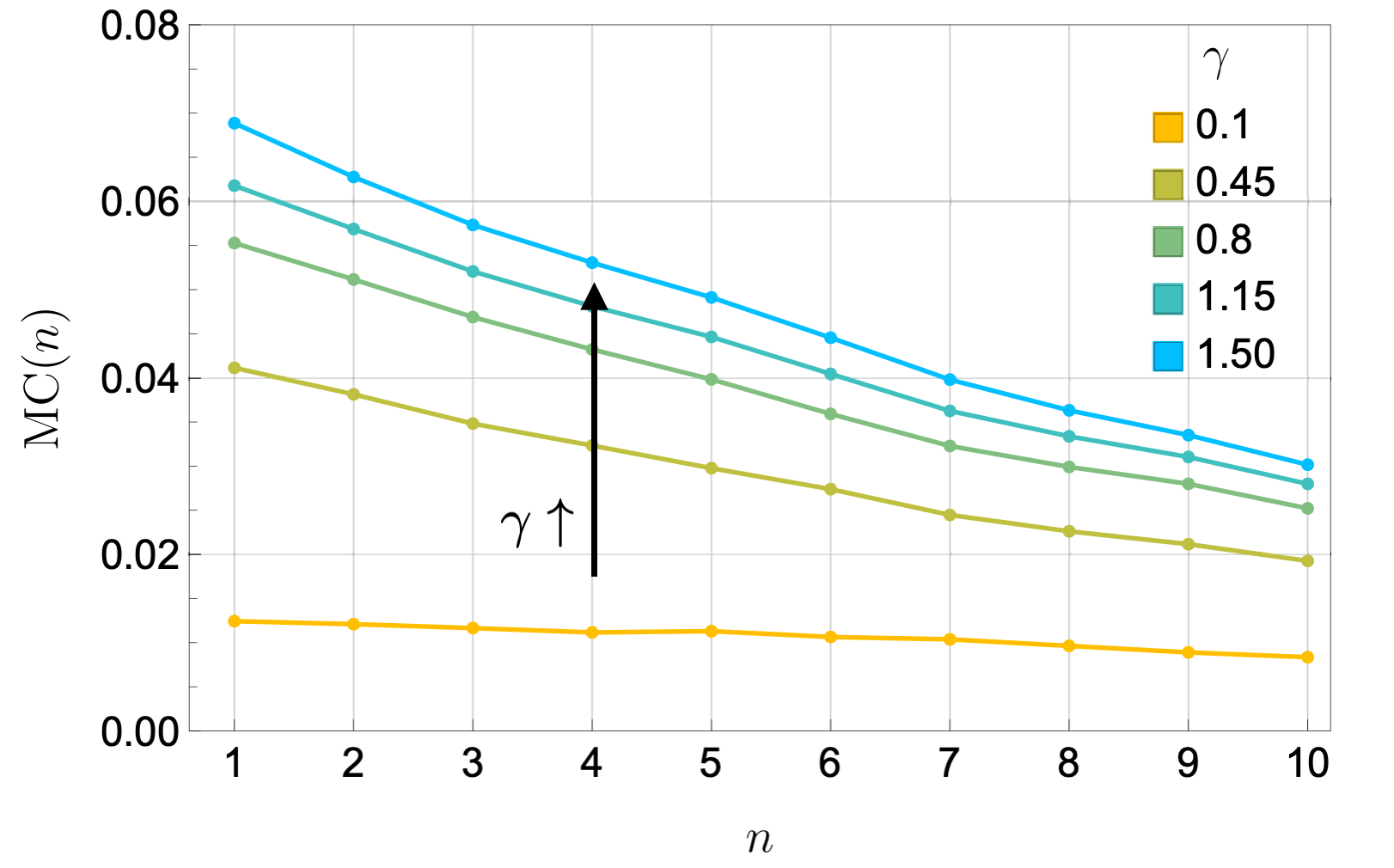}
\caption{Linear scale:  \(\text{MC}(n)\) vs \(n\).}
\end{subfigure} &
\begin{subfigure}{0.45\textwidth}
\includegraphics[keepaspectratio, width = \textwidth]{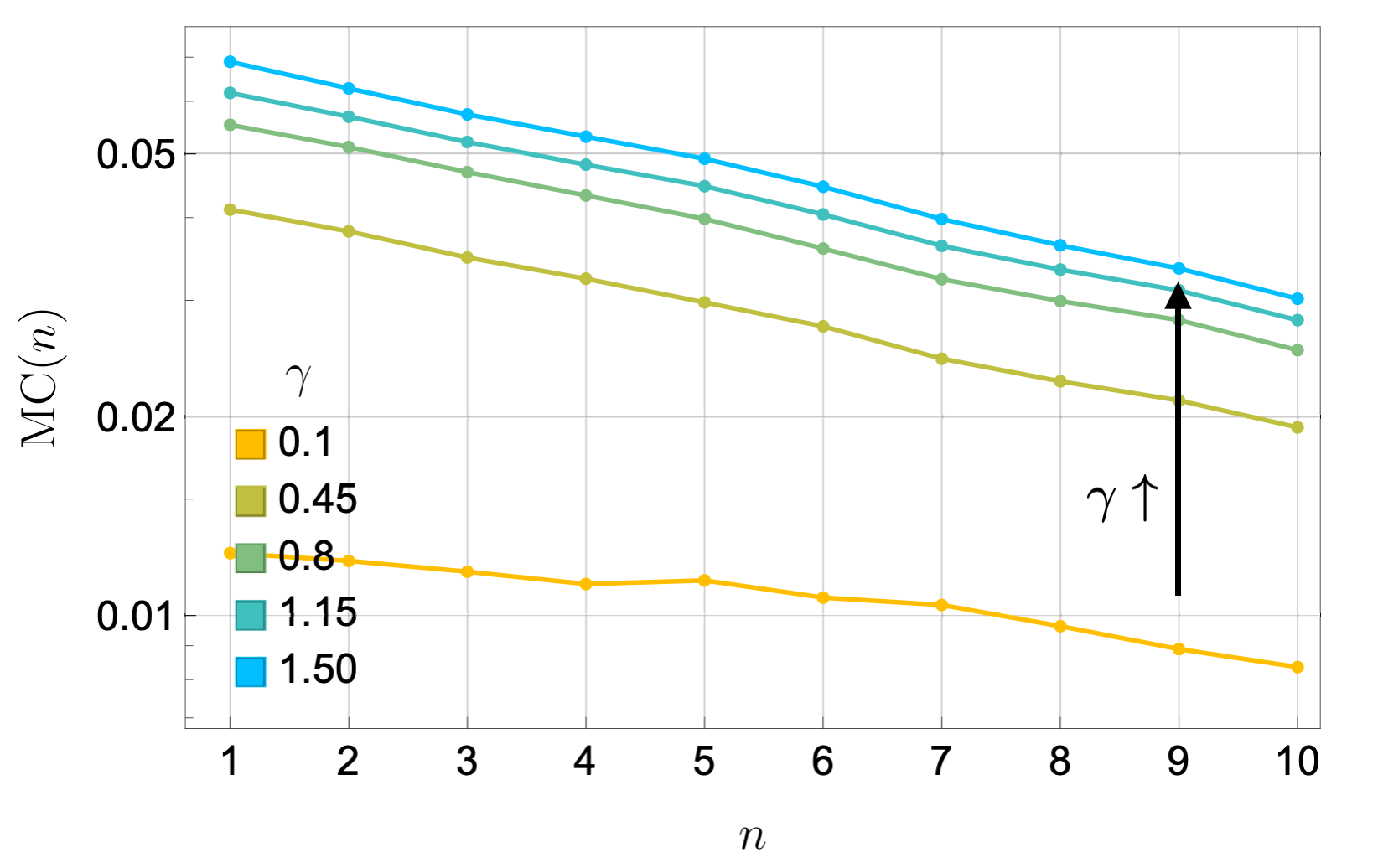}
\caption{Log scale: \(\text{MC}(n) \sim \exp(-\Gamma n)\).}
\end{subfigure}
\end{tabular}
\caption{Memory capacity at the critical coupling $J = |\Delta| = 2$ as \(\gamma\) increases. (a) Average memory capacity over 100 input realizations. (b) Log-scale plot of (a), illustrating exponential decay of memory capacity with delay \(n\). The exponential decay rate of the two-time correlation in the memory capacity, \(\Gamma\), remains approximately constant as \(\gamma\) increases, indicating that dissipation uniformly governs the loss of correlations across different dissipation rates. Differences in memory capacity primarily arise from the proportionality factor, with larger \(\gamma\) leading to a smaller variance of the output prediction in the denominator of Eq.~\eqref{eq:MC}, as the output rapidly stabilizes to the steady state. This rapid stabilization corresponds to highly redundant encoding, and in turn enhances total memory capacity for memorizing uniformly random input time series. Notably, in this redundant coding regime, highly dissipative dynamics improves the quantum reservoir's memory capacity.}
\label{fig:MCvarygamma}
\end{center}
\end{figure}

Figure~\ref{fig:MCvarygamma} illustrates how dissipation (\(\gamma\)) affects memory capacity at the critical coupling \(J = |\Delta| = 2\). As \(\gamma\) increases, the memory capacity can be attributed to the enhanced stability (more rapid relaxation towards) in the reservoir's steady-state dynamics. Dissipation suppresses oscillatory behavior and stabilizes the reservoir's response. This stabilization corresponds to a regime of highly redundant encoding, where information is stored near the steady state across somewhat identical subsystems.

Interestingly, as shown in Fig. \ref{fig:MCvaryJ} while higher \(\gamma\) leads to improved total memory capacity, the decay rate of memory capacity in this redundant regime, \(MC(n) \sim \exp(-\Gamma n)\), remains approximately constant across different dissipation rates. This suggests that dissipation uniformly governs the loss of correlations over time. Differences in memory capacity at high \(\gamma\) primarily arise from the proportionality factor in Eq.~\eqref{eq:MC}. As dissipation increases, the variance of the reservoir's output prediction decreases, reflecting rapid stabilization to the steady state. This reduced variance amplifies the initial memory capacity but does not alter the exponential decay rate of correlations.

These results highlight a subtle interplay between dissipation, memory capacity, and encoding modes. Near criticality at \(J = |\Delta|\), the system exhibits synergistic behavior, where collective dynamical response dominates, enabling the reservoir to respond sensitively to recent input signals. This synergistic encoding boosts short-term memory retention but comes at the expense of long-term storage, as the system's ability to retain correlations with far past inputs diminishes rapidly due to sensitivity to perturbation. 

At higher dissipation rates \(\gamma\), the system tends to encode inputs more redundantly and relaxes more quickly to a steady state. Interestingly, as illustrated in Figure~\ref{fig:MCvarygamma}, both short- and long-term memory capacities, at criticality, increase with \(\gamma\) for a simple uniformly random input memorization task, even though one might initially expect a trade-off. The underlying reason could be that the system has not yet fully achieved the fading memory regime, so further investigations, potentially through more comprehensive benchmarks such as IPC analysis, are needed to confirm how dissipation precisely balances short-term responsiveness and long-term stability right at a critical point.

Also from the perspective of quantum reservoir design, dissipation plays a critical role in ensuring the reservoir exhibits the fading memory property, where the influence of past inputs gradually diminishes, which is a necessary condition for designing operational reservoir computing \cite{sannia2024dissipation, rodrigo_QRCfinite_PRE}.

\section{Conclusion and Outlook}\label{sec:conclusion}

In this work, we explored coupled Kerr-nonlinear oscillators as a model open quantum system for studying modes of information processing in a small quantum reservoir computing (QRC) platform. By employing Partial Information Decomposition (PID) and analyzing memory retention tasks, we investigated how the interplay of dynamical instability and dissipation governs the encoding of input information into redundant or synergistic modes. These encoding schemes play a crucial role in determining the reservoir's performance in processing and retaining temporal data.

Our findings reveal several key insights. First, near the critical coupling strength \(J = |\Delta|\), the system transitions from predominantly redundant to synergistic encoding. This transition is driven by the dynamics of coherent oscillation modes that dominate as slow modes (soft modes) become overdamped. These collective dynamics enable the reservoir to process information synergistically, boosting short-term memory retention. Importantly, this synergy is robust across different input signals, including telegraph processes and uniform uncorrelated noise, suggesting that the observed transition is an intrinsic feature of the system’s bifurcation near \(J = |\Delta|\). This dynamic instability was elucidated through the non-equilibrium (Keldysh) field-theoretic analysis, which highlighted how the disappearance of soft modes amplifies the dominance of fast, coherent modes.

Dissipation (\(\gamma\)) plays an important role in shaping information encoding and memory capacity for our reservoir near criticality. At low \(\gamma\), synergistic encoding enables collective processing and enhances short-term memory retention, as the reservoir's dynamics are sensitive to recent input signals. However, as \(\gamma\) increases, dissipation suppresses coherent dynamics, rapidly driving the system toward redundant encoding. In this regime, large dissipation stabilizes encoding by enforcing redundant representations near the steady state, enabling the reservoir to retain information about far-past inputs at the expense of sensitivity to recent input information.

The connection between encoding modes and memory capacity reveals a trade-off: synergistic encoding favors short-term memory retention but is less suited for long-term storage due to its sensitivity to perturbations. Redundant encoding, on the other hand, sacrifices responsiveness to recent inputs but improves long-term retention. From a quantum reservoir design perspective, dissipation ensures the reservoir exhibits the fading memory property, necessary for effective reservoir design \cite{rodrigo_QRCfinite_PRE, rodrigo_inputdepqrc2024}. By carefully tuning dissipation, one can hope to optimize the balance between short-term responsiveness and long-term stability and tailor the reservoir to specific computational tasks, aligning with recent work on engineered dissipation as computational resources in quantum systems \cite{sannia2024dissipation, verstraete2009quantum, Fry_noiseqrc_2023}.  

Coupled Kerr-nonlinear oscillators provide a minimal yet insightful testbed for analyzing transitions between encoding modes driven by dissipation and dynamical instability. Extending this framework to other quantum systems with dynamical phase transitions, such as Bose-Hubbard models \cite{alaeian2021noise}, driven-dissipative platforms \cite{leghtas2015confining}, and spin systems \cite{martinez2021dynamical, govia2021quantum}, could deepen our understanding of how dissipation and instability shape encoding dynamics in systems with more complex phase spaces.  Another interesting direction is to develop a rigorous definition of quantum synergy. Although this work uses partial information decomposition (PID) to analyze classical information derived from quantum observables, incorporating quantum correlations could provide a more comprehensive view of how coherence and other quantum effects either enhance or constrain information encoding, in line with \cite{govia2021quantum,motamedi2023correlations,gotting2023exploring}.  Such development would refine our understanding of synergy and redundancy in small quantum systems from a more quantum information theoretic viewpoint. 

In practice, realistic quantum devices inevitably face various noise sources and control imperfections. While these perturbations may shift the crossover point between redundant and synergistic encoding or relocate the system’s critical point, our results show that this crossover in the modes of encoding persists over a wide parameter range. With advanced experimental techniques, such as dynamical decoupling and error mitigation in quantum optics, we expect that the essential physics behind this crossover can still be realized in real experimental systems.

Lastly, it is important to recognize that the results presented here, like much of the prior work, assume perfect readout without the presence of shot noise. As system sizes scale, however, readout processes may suffer from exponential concentration phenomena, requiring exponentially many measurement shots to estimate input-dependent readouts accurately, as discussed in \cite{xiong2024QELM}. This limitation presents a significant barrier to scalability. In larger quantum reservoir systems, future studies must address how synergy and redundancy behave under realistic measurement constraints. Incorporating the effects of finite measurement precision into the framework of encoding dynamics could lead to more scalable and experimentally feasible designs for quantum reservoirs.

\section*{Acknowledgement}
This research has received funding support from the NSRF via the Program Management Unit for Human Resources \& Institutional Development, Research and Innovation [grant number B13F660057]. TC acknowledges insightful discussions with Wave Ngampruetikorn on synergistic and redundant coding. 
\appendix
\section{Partial Information Decomposition (PID)}\label{sec:PID}

In classical (Shannon) information theory, we run into conceptual difficulties as soon as there are more than two random variables to handle, because there is no single, universally accepted way to break down the total information among multiple variables following Shannon's original recipe \cite{shannon1948mathematical}. Quantifying the amount of information that is shared (redundant), exclusive (unique), or complementary (synergistic) among three or more variables is highly nontrivial and remains an active area of research, see for example \cite{bell2003co, james2011anatomy, harder2013bivariate, olbrich2015information, rosas2019quantifying, gutknecht2023babel}.
In this work, we adopt one particular framework that meets three criteria: (1) It has a relatively straightforward formalism.
(2) It captures whether and how our system is redundant or synergistic.
(3) It does not exhibit major interpretational drawbacks for our purposes.

This framework, called \textit{partial information decomposition} (PID) \cite{williams2010nonnegative}, attempts to disentangle the multivariate information into non-overlapping, non-negative parts with clear interpretations. Concretely, consider a set of $n$ random variables $\{S, R_1, R_2, \ldots, R_{n-1} \}$, where $S$ (the \textit{target}) is the variable whose information we want to capture, and $\mathbf{R} = \{R_1, \ldots, R_{n-1}\}$ (the \textit{sources}) is the combined set of variables providing that information. The total (multivariate) mutual information between the source and the target is (for discrete distributions):
\begin{equation*}
I(S:\mathbf{R}) 
\;=\; 
\sum_{s,r_1,\ldots,r_{n-1}}
P_{(S,\mathbf{R})}(s,r_1,\ldots, r_{n-1}) 
\log\!\Biggl( \frac{P_{(S,\mathbf{R})}(s, r_1,\ldots, r_{n-1})}{P_S(s)\,P_{\mathbf{R}}(r_1,\ldots, r_{n-1})} \Biggr).
\end{equation*}

\noindent
As outlined in the original PID paper \cite{williams2010nonnegative}, this methodology can in principle be generalized to any number of variables in terms of PID lattice, but its complexity increases rapidly once $n > 3$. (Further developments can be found in \cite{bertschinger2013shared, banerjee2015synergy, wibral2017partial, makkeh2021introducing, schick2021partial}.) In this work, we focus on the simplest nontrivial case $n = 3$, for which the PID formalism is most concretely developed and comparatively well-understood.

\paragraph{PID for three variables.} 
Let $X,Y,Z$ be three random variables, and consider $I(X : (Y,Z))$, the mutual information between $X$ and the pair $(Y,Z)$. The three-variable PID proposes the following decomposition of $I(X : (Y,Z))$ into four distinct parts:
\begin{equation}
    I(X:(Y,Z)) \;=\;
    \text{Rdn}(X:Y;Z) 
    \;+\;
    \text{Syn}(X:Y;Z)
    \;+\;
    \text{Unq}(X:Y\setminus Z) 
    \;+\;
    \text{Unq}(X:Z\setminus Y),
    \label{eqn:PID1}
\end{equation}
along with corresponding decompositions of the pairwise mutual informations:
\begin{align}
    I(X:Y) \;=\;& \text{Rdn}(X:Y;Z) \;+\; \text{Unq}(X:Y\setminus Z), 
    \label{eqn:PID2} \\
    I(X:Z) \;=\;& \text{Rdn}(X:Y;Z) \;+\; \text{Unq}(X:Z\setminus Y). 
    \label{eqn:PID3}
\end{align}
Here, the four partial information (PI) terms have the following interpretations.
\begin{itemize}
    \item $\text{Rdn}(X:Y;Z)$ (\emph{redundancy/shared information}): the amount of information about $X$ that is found \emph{in common} in $Y$ and $Z$.
    \item $\text{Syn}(X:Y;Z)$ (\emph{synergy/complementary information}): the information about $X$ that is only accessible when considering $Y$ and $Z$ \emph{jointly}, i.e.\ the “whole is greater than the sum of its parts,” which is typically a signature of cooperation of constituents in complex systems.
    \item $\text{Unq}(X:Y\setminus Z), \text{Unq}(X:Z\setminus Y)$ (\emph{unique information}): the information about $X$ that can be acquired solely from $Y$ (or $Z$) and not from $Z$ (or $Y$).  
\end{itemize}

\noindent
Because there are four unknown PI quantities but only three equations \eqref{eqn:PID1}--\eqref{eqn:PID3}, one cannot solve for the PIs without additional theoretical constraints. Different PID axioms or definitions fix one of these quantities first (e.g., by prescribing a formula or an inequality), allowing the remaining quantities to be determined from the joint distribution $P(X,Y,Z)$. Approaches in the literature include:
(1) Defining redundancy first \cite{williams2010nonnegative, harder2013bivariate, rauh2017extractable, makkeh2021introducing}.
(2) Defining synergy first \cite{griffith2014quantifying, olbrich2015information}.
(3) Defining unique information first \cite{bertschinger2014quantifying}.

\paragraph{Co-information.}
From \eqref{eqn:PID1}--\eqref{eqn:PID3}, one obtains the following notable combination of mutual informations
\begin{equation}
    I(X:Y) + I(X:Z) - I\bigl(X:(Y,Z)\bigr)
    \;=\;
    \text{Rdn}(X:Y;Z) \;-\; \text{Syn}(X:Y;Z).
    \label{eqn:co-info_intro}
\end{equation}
The left-hand side is often called the \emph{co-information} \cite{bell2003co} (or equivalently, \emph{interaction information} \cite{mcgill1954multivariate} or sometimes just the \emph{mutual information} \cite{yeung1991new}):
\begin{equation}
    \text{CoI}(X;Y;Z)
    \;=\;
    I(X:Y) + I(X:Z) \;-\; I\bigl(X:(Y,Z)\bigr).
    \label{eqn:co-info}
\end{equation}
Although this is the simplest and most direct extension of mutual information to three variables, it has two important drawbacks. First, it can take positive or negative values (making some interpretations more subtle than the nonnegative mutual information).
Secondly, it can not really distinguish redundancy and synergy, because the right-hand side of \eqref{eqn:co-info_intro} just reflect their differences.

Nevertheless, for three-variable systems, $\text{CoI}$ can \emph{sign} whether redundancy or synergy dominates. Even though the co-information cannot quantify the amount of redundancy and synergy separately, it can describe whether the system is in \textit{redundant coding} or \textit{synergistic coding} regime \cite{schneidman2003synergy}. When $\text{CoI} > 0$ the system is said to be \textit{redundancy-dominated} and when $\text{CoI} < 0$ the system is said to be \textit{synergy-dominated} \cite{schneidman2003network, rosas2019quantifying}. This measure also plays a role in certain PID definitions and algorithms \cite{bertschinger2014quantifying}.


\section{Explicit Examples for PID Calculation} \label{app:PID}

In this appendix, we provide simple but useful examples to show how one can compute Partial Information Decomposition (PID) terms analytically under the BROJA approach \cite{bertschinger2014quantifying}.  The BROJA method obtains each partial information by solving a constrained optimization problem over all joint distributions \(Q(X,Y,Z)\) consistent with certain marginal constraints derived from the true distribution \(P(X,Y,Z)\).  

\paragraph{General Framework.}

Let \(X,Y,Z\) be three discrete random variables with a true joint distribution \(P(X,Y,Z)\).  The BROJA definitions of the four partial information terms (\(\text{Unq}, \text{Rdn}, \text{Syn}\)) are as follows \cite{bertschinger2014quantifying}:

\begin{equation}
\begin{aligned}
\text{Unq}(X:Y\setminus Z) 
&= 
\min_{Q\in \Delta_P} \, I_Q(X:Y \mid Z),  
\\[6pt]
\text{Unq}(X:Z\setminus Y) 
&= 
\min_{Q \in \Delta_P} \, I_Q(X:Z \mid Y), 
\\[6pt]
\text{Rdn}(X:Y;Z) 
&= 
\max_{Q \in \Delta_P} \, \text{CoI}_Q(X;Y;Z),  
\\[6pt]
\text{Syn}(X:Y;Z) 
&= 
I(X:(Y,Z)) 
\;-\;
\min_{Q \in \Delta_P} \, I_Q(X:(Y,Z)),
\label{eqn:BROJA}
\end{aligned}
\end{equation}
where \(\text{CoI}_Q(X;Y;Z)\) is the co-information under the joint distribution \(Q\), and 
\[
I(X:(Y,Z)) \;=\; I_P(X:(Y,Z))
\]
is the mutual information computed from the {\it actual} distribution \(P\).  Any subscripted quantity such as \(I_Q(\cdot)\) is computed from a {\it candidate} joint distribution \(Q\in\Delta_P\), not from the true \(P\).  

\paragraph{Definition of \(\Delta_P\).}
The space \(\Delta_P\subseteq \Delta\) consists of all joint distributions \(Q(X,Y,Z)\) whose \((X,Y)\) and \((X,Z)\) marginals match those of \(P\).  Formally, if \(\Delta\) is the set of all distributions on \(\mathcal{X}\times \mathcal{Y}\times \mathcal{Z}\), then
\begin{equation}
    \Delta_P 
    \;=\; 
    \Bigl\{ Q \in \Delta 
    \;\Big|\; 
        Q(x,y) = P(x,y)
        \,\text{ and }\,
        Q(x,z) = P(x,z) 
        \;\Bigr\}, 
\end{equation}
for all \((x,y,z)\in \mathcal{X}\times \mathcal{Y}\times \mathcal{Z}\).  In other words, we fix the two-dimensional marginals of \((X,Y)\) and \((X,Z)\) to match the true data, but allow \(Q(Y,Z|X)\) to vary.

\paragraph{Foliation into Slices.} 
To handle the high dimensionality of \(\Delta_P\), a convenient approach parameterizes \(\Delta_P\) via a {\it foliation}, see appendix A. of \cite{bertschinger2014quantifying}.  Concretely, for each \(x\) with \(P(x)>0\), we define
\begin{equation}
\Delta_{P,x} 
\;=\; 
\Bigl\{ \,
\tilde{Q}(Y,Z)\in \Delta(\mathcal{Y}\times\mathcal{Z}) 
\;\Big|\; 
\tilde{Q}(y) = P(y|x), \ 
\tilde{Q}(z) = P(z|x) 
\,\Bigr\}.
\label{eqn:P_slice}
\end{equation}
A joint distribution \(Q\in \Delta_P\) can then be specified by choosing, for each \(x\), a \(\tilde{Q}(Y,Z)\in \Delta_{P,x}\), and combining via
\begin{equation}
Q(x,y,z) \;=\; 
P(x)\,\tilde{Q}_x(y,z) = P(x)Q(y,z|x).
\label{eqn:Q_combine}
\end{equation}
Hence, \(\Delta_P\) can be viewed as the product of slice spaces \(\Delta_{P,x}\) for all \(x\).  Once a suitable parameterization of these slices is chosen, the optimization in \eqref{eqn:BROJA} becomes a finite-dimensional convex optimization problem. Next, we show how the procedure is performed in the simple cases following \cite{bertschinger2014quantifying}.

\paragraph{Example 1: AND Gate.}

Consider three binary variables \(\mathcal{X}=\mathcal{Y}=\mathcal{Z}=\{0,1\}\) with
\[
X = (Y \,\mathrm{AND}\, Z)
\]
and \(Y, Z\) i.i.d.\ \(\mathrm{Bernoulli}(1/2)\).  The {\it true} joint distribution \(P\) is uniform on the events \((X,Y,Z)\in\{(0,0,0),(0,0,1),(0,1,0),(1,1,1)\}\) with probability \(1/4\) each, and zero otherwise.  From this \(P\), one derives the marginals \(P(X,Y)\) and \(P(X,Z)\).  

\begin{figure}[h!]
    \centering
    \includegraphics[width=0.45\textwidth]{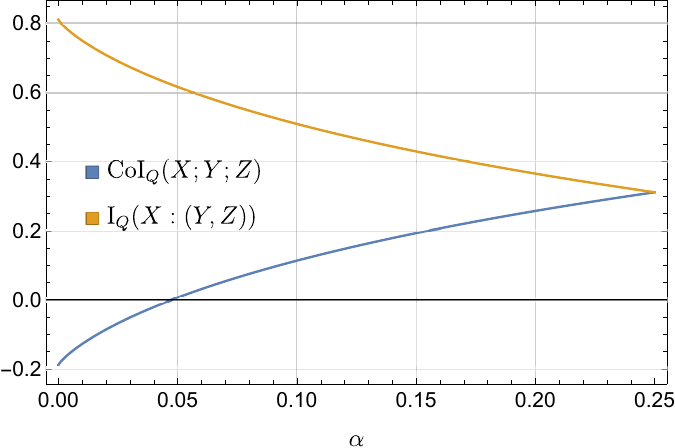}
    \caption{\small Co-information \(\text{CoI}_{Q_\alpha}(X;Y;Z)\) and mutual information \(I_{Q_\alpha}(X:(Y,Z))\) for the AND-gate example, plotted as functions of \(\alpha\).  The optimum for redundancy (resp.\ synergy) occurs at \(\alpha=\tfrac{1}{4}\).}
    \label{fig:AND_example}
\end{figure}

\paragraph{Parametrizing \(\Delta_P\).} 
Applying the slice formalism, one obtains:
\begin{itemize}
    \item For \(x=0\) (which has \(P(X=0)=3/4\)), the slice \(\Delta_{P,0}\) is a 1-parameter family 
    \[
        \tilde{Q}_0(0,0) 
        = 
        \tfrac{1}{3} + \alpha', \quad
        \tilde{Q}_0(0,1) 
        = 
        \tfrac{1}{3}-\alpha', 
        \quad
        \tilde{Q}_0(1,0) 
        = 
        \tfrac{1}{3}-\alpha', 
        \quad
        \tilde{Q}_0(1,1) 
        = 
        \alpha',
    \]
    with \(0\le \alpha' \le 1/3\). 
    \item For \(x=1\) (which has \(P(X=1)=1/4\)), the slice \(\Delta_{P,1}\) is trivial because the only consistent distribution is \(\tilde{Q}_1(1,1)=1\). 
\end{itemize}
Combining \(\tilde{Q}_0\) and \(\tilde{Q}_1\) as in \eqref{eqn:Q_combine} and reparameterizing yields the 1-parameter family
\[
Q_{\alpha}(x,y,z) 
=\;
\begin{cases}
\frac{1}{4} + \alpha, & (x,y,z) = (0,0,0),\\
\frac{1}{4} - \alpha, & (x,y,z) = (0,0,1), (0,1,0),\\
\alpha, & (x,y,z) = (0,1,1),\\
\frac{1}{4}, & (x,y,z) = (1,1,1),
\end{cases}
\quad
0 \,\le\, \alpha \,\le\, \frac{1}{4}.
\]
Under each candidate \(Q_{\alpha}\), one can compute \(\text{CoI}_{Q_\alpha}(X;Y;Z)\) and \(I_{Q_\alpha}(X:(Y,Z))\), and thus solve the optimization problems over \(\alpha\) in \eqref{eqn:BROJA}.  Figure~\ref{fig:AND_example} shows \(\text{CoI}_{Q_\alpha}\) and \(I_{Q_\alpha}\) vs.\ \(\alpha\).

Solving these optimizations shows that \(\alpha=\tfrac{1}{4}\) is the critical point.  Hence, each partial information is
\begin{align*}
\text{Unq}(X:Y\setminus Z) &= \text{Unq}(X:Z\setminus Y) = 0    \\
\text{Rdn}(X:Y;Z) &= \frac{3}{4}\log{\frac{4}{3}} = 0.311\log{2} = 0.311 \:\text{bit} \\
\text{Syn}(X:Y;Z) &= \frac{1}{2}\log{2} = 0.5 \:\text{bit} 
\end{align*}

\paragraph{Example 2: XOR Gate.}

Next, consider the binary XOR relation
\[
X = (Y \,\mathrm{XOR}\, Z), 
\]
where \(Y, Z\) are again i.i.d.\ \(\mathrm{Bernoulli}(1/2)\).  The true distribution \(P\) is uniform on \(\{(0,0,0), (1,0,1), (1,1,0), (0,1,1)\}\). One again sets up the slices \(\Delta_{P,0}\) and \(\Delta_{P,1}\), each giving a family of distributions parameterized by \(\alpha'\) and \(\beta'\). 

\begin{figure}[h!]
\centering
\begin{tabular}{cc}
\includegraphics[width=0.4\textwidth]{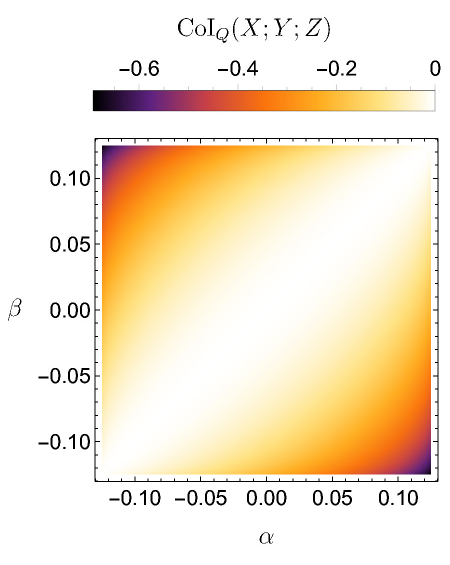}
& 
\includegraphics[width=0.4\textwidth]{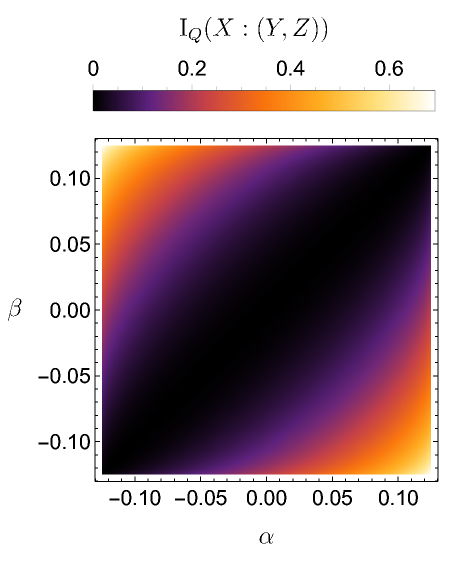}
\end{tabular}
\caption{\small 
(Left) \(\text{CoI}_{Q_{\alpha,\beta}}(X;Y;Z)\); 
(Right) \(I_{Q_{\alpha,\beta}}(X:(Y,Z))\); 
for the XOR-gate example.  
The optimum occurs at \(\alpha=0,\beta=0\).}
\label{fig:XOR_example}
\end{figure}

\begin{itemize}
\item $\Delta_{P,0}$: the most general distribution that satisfies the constraint \eqref{eqn:P_slice} is
    \begin{equation*}
        \tilde{Q}_0( y, z) = \left\{
            \begin{array}{ll}
                \alpha' &, \quad(y,z) = (0,0) \\
                \frac{1}{2} - \alpha' &, \quad(y,z) = (0,1) \\
                \frac{1}{2} - \alpha' &, \quad(y,z) = (1,0) \\
                \alpha' &, \quad(y,z) = (1,1)
            \end{array}
        \right.
    \end{equation*}
with $0 \leq \alpha' \leq \frac{1}{2}$.
\item $\Delta_{P,1}$: since the distribution for \text{XOR} operation is symmetric under swapping $Y$ and $Z$, then it turns out that  
    \begin{equation*}
        \tilde{Q}_1(y, z) = \left\{
            \begin{array}{ll}
                \beta' &, \quad(y,z) = (0,0) \\
                \frac{1}{2} - \beta' &, \quad(y,z) = (0,1) \\
                \frac{1}{2} - \beta' &, \quad(y,z) = (1,0) \\
                \beta' &, \quad(y,z) = (1,1)
            \end{array}
        \right.
    \end{equation*}
with another free parameter, $0 \leq \beta' \leq \frac{1}{2}$.
\end{itemize}
Combining them with \(P(X=0)=P(X=1)=\tfrac12\) yields a two-parameter family 
\begin{equation*}
    Q_{\alpha,\beta}(x, y, z) = \left\{
        \begin{array}{ll}
            \frac{1}{8} + \alpha &, \quad(x,y,z) = (0,0,0), (0,1,1) \\
            \frac{1}{8} - \alpha &, \quad(x,y,z) = (0,0,1), (0,1,0) \\
            \frac{1}{8} + \beta  &, \quad(x,y,z) = (1,0,0), (1,1,1) \\
            \frac{1}{8} - \beta  &, \quad(x,y,z) = (1,0,1), (1,1,0) \\
        \end{array}
    \right.
\end{equation*}
where $-\frac{1}{8} \leq \alpha, \beta \leq \frac{1}{8}$, governing all \(Q\in \Delta_P\).  As before, one computes \(\text{CoI}_{Q_{\alpha,\beta}}(X;Y;Z)\) and \(I_{Q_{\alpha,\beta}}(X:(Y,Z))\) to solve the optimization in \eqref{eqn:BROJA}.  Figure~\ref{fig:XOR_example} shows the surfaces of \(\text{CoI}\) and \(I\) vs.\ \(\alpha,\beta\).

The optimum occurs at \((\alpha,\beta)=(0,0)\), giving
\begin{align*}
\text{Unq}(X:Y\setminus Z) &= \text{Unq}(X:Z\setminus Y) = 0    \\
\text{Rdn}(X:Y;Z) &= 0 \\
\text{Syn}(X:Y;Z) &= \log{2} = 1 \:\text{bit} .
\end{align*}
 
This perfectly aligns with the well-known fact that XOR is {\it synergistic}. Consider $X = Y \:\text{XOR} \:Z$ and we happen to have only the knowledge of $Y$, either we know that $Y = 0$ or $Y = 1$. Without the knowledge of $Z$, we cannot infer any useful information about the value of $X$ at all, the probability $P(X|Y)$ is always equal to $1/2$ for any $(x,y)$. Only when both $Y$ and $Z$ are presented together do we know exactly what the value of $X$ should be and it is \textit{exactly} 1 bit of information (i.e. one yes/no question) that we need to know in order to eliminate all uncertainty about the value of $X$.

These two classic gates (AND and XOR) show how the BROJA optimization can be computed analytically in simple discrete cases.  In practice, for larger or more complex systems, numerical methods are necessary, but the underlying principle is the same. One restricts to \(\Delta_P\) to preserve certain marginals, and then solves the convex optimization problems in \eqref{eqn:BROJA}.

\section{Keldysh Action}
\label{app:Keldysh}

In this appendix, we outline how to obtain the Keldysh action for the coupled Kerr oscillators described by Eqs.~\eqref{eqn:Hamiltonian} and \eqref{eqn:Lindblad}, and how the system’s linear response to external perturbation can be obtained from the \emph{retarded} Green’s function. Readers seeking broader context on Keldysh formalism may consult \cite{dalla2013keldysh, sieberer2016keldysh, kamenev2023field}.

From the Linblad equation.~\eqref{eqn:Hamiltonian}, one can reformulate this open-system evolution in a path-integral language by introducing ``classical'' fields $\alpha_{j,c}$ and ``quantum'' fields $\alpha_{j,q}$, capturing forward and backward time contours, respectively \cite{kamenev2023field}. After performing the usual Keldysh rotation, the action $S_K$ takes the following form
\begin{align}
S_K 
\,=\,
\int dt
\bigl[\,\,&\sum_{j=1,2}
\Bigl(
\alpha_{j,c}^*\,i\partial_t\,\alpha_{j,q}
+\,
\alpha_{j,q}^*\,i\partial_t\,\alpha_{j,c}
-\,
\Delta_j\bigl(\alpha_{j,c}^*\alpha_{j,q} 
+ 
\alpha_{j,q}^*\alpha_{j,c}\bigr)
\Bigr)
\nonumber\\
&-\,
J\bigl(
\alpha_{1,q}^*\alpha_{2,c}
+ 
\alpha_{1,c}^*\alpha_{2,q}
+
\alpha_{2,q}^*\alpha_{1,c}
+ 
\alpha_{2,c}^*\alpha_{1,q}
\bigr)
\nonumber\\
&-\,
\tfrac{1}{2}\sum_{j=1,2}U_j
\bigl(
|\alpha_{j,c}|^2 + |\alpha_{j,q}|^2
\bigr)
\bigl(
\alpha_{j,c}^*\alpha_{j,q}
+ 
\alpha_{j,c}\alpha_{j,q}^*
\bigr)
\nonumber\\
&-\,
\sqrt{2}\,F(t)\,\sum_{j=1,2}
\bigl(\alpha_{j,q}^* + \alpha_{j,q}\bigr)
+\,i\sum_{j=1,2}\gamma_j
\bigl(
2|\alpha_{j,q}|^2
+
\alpha_{j,c}\alpha_{j,q}^*
-\,
\alpha_{j,c}^*\alpha_{j,q}
\bigr)\bigr].
 \label{eqn:action}
\end{align}
Here, $\Delta_j$ is the detuning frequency, $U_j$ the Kerr nonlinearity, $J$ the coupling rate, and $F(t)$ an external drive. Varying $S_K$ with respect to $\alpha_{j,c}^*$ and $\alpha_{j,q}^*$ yields the semiclassical equations of motion that incorporate both Hamiltonian and dissipative dynamics.

\paragraph{Mean-field equations and the effective potential.}

Mean-field or semiclassical (or saddle-point) dynamics are found by setting
$
\frac{\delta S_K}{\delta \alpha_{j,c}}
\,=\,
\frac{\delta S_K}{\delta \alpha_{j,q}}
\,=\,
0
$
and similarly for complex conjugates. One obtains $\alpha_{j,q}=0$ as a trivial solution \cite{kamenev2023field}, and the mean-field dynamics for $\alpha_{j,c}$ then follow from
$\frac{\delta S_K}{\delta \alpha_{j,q}^*}\Big|_{\alpha_{j,q}=0}\,=\,0$ giving
\begin{equation}\label{eqn:classicalEOM}
\partial_t \alpha_{j,c}
\,=\,
-\bigl(\gamma_j + i\Delta_j\bigr)\alpha_{j,c}
\;-\;
iJ\,\alpha_{j',c}
\;-\;
\tfrac{i}{2}\,U_j\,\alpha_{j,c}\,|\alpha_{j,c}|^2
\;-\;
i\sqrt{2}\,F(t),
\end{equation}
where $j' \neq j$.
This can be recast as a potential dynamics of the form,
\[
i\,\partial_t\alpha_{j,c}
\,=\,
\partial_{\alpha_{j,c}^*}V(\vec{\alpha}_c,\vec{\alpha}_c^*)
\;-\;
i\,\gamma_j\,\alpha_{j,c}
\;+\;
\sqrt{2}\,F(t),
\]
with the {\it effective potential} 
\[
V(\vec{\alpha}_c,\vec{\alpha}_c^*)
\,=\,
\sum_{j=1,2}
\bigl(
\Delta_j\,|\alpha_{j,c}|^2
+
\tfrac{1}{4}\,U_j\,|\alpha_{j,c}|^4
\bigr)
\;+\;
J\bigl(\alpha_{1,c}\,\alpha_{2,c}^*
+\,
\alpha_{2,c}\,\alpha_{1,c}^*\bigr),
\]
which is the potential landscape discussed in Sec.~\ref{sec:slow_modes}.

\paragraph{Fluctuations and the inverse Green’s function.}
To analyze the system's linear response to small perturbation about a mean-field or saddle-point solution, we expand the Keldysh action to second order in the fluctuating fields $\delta\alpha_{j,c/q}$. Namely, we set
\[
\alpha_{j,c/q}(t)
\,\to\,
\alpha_{j,c/q}(t)
\;+\;
\delta\alpha_{j,c/q}(t),
\]
expand $S_K$ in \eqref{eqn:action} to {\it quadratic} order in $\delta\alpha_{j,c/q}$, and then consider the Fourier decomposition of the fluctuations 
\[
\delta\alpha_{c/q}(t)
\,=\,
\frac{1}{\sqrt{2}}\int\!\!
\frac{d\omega}{2\pi}
\,\bigl[
\delta\alpha_{c/q}(\omega)\,e^{-i\omega t}
\,+\,
\delta\alpha_{c/q}(-\omega)\,e^{+\,i\omega t}
\bigr].
\]
In block-matrix form, the resulting {\it quadratic} action reads
\begin{equation}\label{eq:quaratic_action}
S_{K,2}
\,=\,
\tfrac{1}{2}\!\int\!\frac{d\omega}{2\pi}\,
\delta\mathbf{\Phi}^\dagger(\omega)
\,
\begin{pmatrix}
0 & [G^A(\omega)]^{-1}
\\[6pt]
[G^R(\omega)]^{-1} & D^K
\end{pmatrix}
\!\delta\mathbf{\Phi}(\omega),
\end{equation}
where $\delta\mathbf{\Phi}(\omega)$ groups the fluctuations $\{\delta\alpha_{j,c/q}(\omega),\,\delta\alpha_{j,c/q}^*(-\omega)\}$.
Explicitly, in the vector component such that \\
\( \delta\mathbf{\Phi}(\omega) = \Big( \delta\alpha_{1,c}(\omega), \delta\alpha_{1,c}^*(-\omega) , \delta\alpha_{2,c}(\omega), \delta\alpha_{2,c}^*(-\omega), \delta\alpha_{1,q}(\omega) , \delta\alpha_{1,q}^*(-\omega) ,  \delta\alpha_{2,q}(\omega), \delta\alpha_{2,q}^*(-\omega) \Big)^T, \)
the inverse of {\it retarded/advanced Green's function} $[G^{R/A}(\omega)]^{-1}$ and the Keldysh component of the inverse Green's function $D^K$ read, respectively, 
\begin{align}
[G^R(\omega)]^{-1} &= \mqty( \omega - g_1 +i\gamma_1 & -\frac{1}{2}U_1\alpha_{1,c}^2 & -J & 0 \\
    -\frac{1}{2}U_1(\alpha_{1,c}^*)^2 & -\omega - g_1 -i\gamma_1  & 0 & -J \\
    -J & 0 & \omega - g_2 +i\gamma_2 & -\frac{1}{2}U_2\alpha_{2,c}^2 \\
    0 & -J & -\frac{1}{2}U_2(\alpha_{2,c}^*)^2 & -\omega - g_2 -i\gamma_2 ) \label{eqn:Greenfunction}, \\
[G^A(\omega)]^{-1} &= [G^R(\omega)^{\dagger}]^{-1},  \  \text{and}\\
D^K &= 2i\,\text{diag}(\gamma_1, \gamma_1, \gamma_2, \gamma_2),
\end{align}
where $g_j = \Delta_j+ U_j|\alpha_{j,c}|^2$ for $j = 1,2$. 

\paragraph{Classical field response.}
We are interested in the fluctuations of the \emph{classical} field variables $\delta\alpha_{j,c}(\omega)$, which is precisely encoded the inverse retarded block, $[G^R(\omega)]^{-1}$ of~\eqref{eq:quaratic_action}, that determines how those fluctuations grow or decay. Let $\delta F(t)$ be a small external drive coupling linearly to the oscillator modes as in~\eqref{eqn:classicalEOM}. At the level of fluctuations, one can write
\[
\delta\alpha_{j,c}(\omega)
=\,
G^R_{j}(\omega)\,\delta F(\omega),
\]
where $G^R_{j}(\omega)$ is the relevant component (or linear combination of components) of the retarded Green’s function. In the time domain,
\[
\delta\alpha_{j,c}(t)
=\,
\int_{-\infty}^{\infty}dt'\,\,
G^R_{j}(t - t')\,\delta F(t'),
\]
with $G^R_{j}(t) = 0$ for $t<0$. This causality condition sets the retarded (not advanced) nature of $G^R$; the system cannot respond \emph{before} the perturbation arises.

\paragraph{Pole Structure and Oscillation vs.\ Decay.}
To analyze the dynamics near a stationary solution, e.g.\ $\alpha_{1,c} = \alpha_{2,c} = 0$, we linearize around that point and compute $G^R(\omega)$ by inverting the block matrix. There, the inverse retarded Green's function of~\eqref{eq:quaratic_action} becomes 
\begin{equation}
[G^R(\omega)]^{-1} = \mqty( \omega -\Delta_1 +i\gamma_1 &0 & -J & 0 \\
    0 & -\omega -\Delta_1 -i\gamma_1  & 0 & -J \\
    -J & 0 & \omega - \Delta_2 +i\gamma_2 & 0 \\
    0 & -J & 0 & -\omega - \Delta_2 -i\gamma_2 ) \label{eqn:Greenfunction}. 
\end{equation}
The poles of $G^R(\omega)$ appear where
$[G^R(\omega)]^{-1}
\,\to\,
0,
$
i.e., where the determinant of the retarded block vanishes. Writing such a pole as $\omega_{\star} = \Omega \pm i\,\Gamma$ clarifies the physical nature of the fluctuation mode:
\begin{itemize}
\item $\Omega = \mathrm{Re}(\omega_{\star})$ is the frequency of oscillation.  
\item $\Gamma = -\,\mathrm{Im}(\omega_{\star})$ is the exponential decay rate if $\Gamma>0$, signifying a stable, dissipative mode.
\end{itemize}

By setting $\Delta_1 = \Delta_2 = \Delta$ and $\gamma_1 = \gamma_2 = \gamma > 0$ as discussed in Sec.~\ref{sec:slow_modes}, one obtains the poles describing the fast and slow relaxation modes in~\eqref{eq:G_retarded_poles}, that is 
\[
\omega_s \;=\; \pm\!\bigl|J - |\Delta|\bigr|\;-\;i\,\gamma,
\quad
\omega_f \;=\; \pm\!\bigl|J + |\Delta|\bigr|\;-\;i\,\gamma.
\]
The real part of a slow-mode pole start to disappear when the effective potential becomes marginally flat (e.g., $J \simeq |\Delta|$). Although one might expect a zero-frequency “soft” oscillation in a conservative dynamics setting, dissipation (encoded in $-\mathrm{i}\gamma$ terms of $[G^R(\omega)]^{-1}$) shifts that would-be neutral oscillatory mode into an overdamped decay channel.

In summary, the Keldysh action formalism provides a powerful lens to derive both the mean-field equations of motion and the fluctuation response in an open quantum system. Retarded Green’s functions, in particular, capture the causality of how a driven perturbation modifies the system at later times, thereby revealing the presence of overdamped or oscillatory collective modes. These results underpin the discussion in Sec.~\ref{sec:slow_modes} on how the disapperance of slow modes leads to synergistic encoding at $J\approx|\Delta|$.


\section{Second-Order Cumulant Expansion}\label{app:cumulant}

In this appendix, we derive the second-order (2\textsuperscript{nd}-order) cumulant expansion used to obtain semiclassical equations of motion for eight complex-valued expectation variables
\[
\bigl\{ 
\langle \hat{a}_{1} \rangle,\,
\langle \hat{a}_{2} \rangle,\,
\langle \hat{n}_{1} \rangle,\,
\langle \hat{n}_{2} \rangle,\,
\langle \hat{a}_{1}^{2}\rangle,\,
\langle \hat{a}_{2}^{2}\rangle,\,
\langle \hat{a}_{1}^{\dagger}\hat{a}_{2}\rangle,\,
\langle \hat{a}_{1}\hat{a}_{2}\rangle
\bigr\}.
\]
This second-order cumulant expansion serves as an interpolation between simpler mean-field dynamics, where second-order cumulants factorize and yield no correlations, and the full quantum description, in which higher-order cumulants can be nonzero when quantum correlations are sufficiently strong.
We follow the standard cumulant-truncation scheme \cite{kubo1962generalized}, imposing that all third- and fourth-order cumulants vanish \cite{sanchez2020cumulant}. That is, we set
\[
\langle \hat{A}\hat{B}\hat{C} \rangle_{C} = 0 
\quad \text{and} \quad 
\langle \hat{A}\hat{B}\hat{C}\hat{D} \rangle_{C} = 0,
\]
where the subscript ``$C$'' denotes the connected (cumulant) part. This approximation can capture second-order correlations while keeping the system of equations tractable.

Below, we provide the resulting coupled ordinary differential equations (ODEs). These govern the dynamics of our reduced set of expectation values under the second-order expansion.  

\begin{align}
\frac{d}{dt}\langle \hat{a}_j \rangle \;=\;& 
-(\gamma_j + i \Delta_j)\langle \hat{a}_j \rangle 
- iU_j \Bigl( 
   \langle \hat{a}_j^{\dagger}\rangle\,\langle \hat{a}_j^2\rangle 
   + 2\,\langle \hat{a}_j\rangle\,\langle \hat{n}_j\rangle 
   - 2\,\langle \hat{a}_j^{\dagger} \rangle\,\langle \hat{a}_j\rangle^2 
\Bigr) 
\nonumber \\
& \quad 
- iJ \Bigl( \delta_{1j}\,\langle \hat{a}_2\rangle + \delta_{2j}\,\langle \hat{a}_1\rangle\Bigr) 
- i\,F(t)\,,
\label{eq:a_j}
\\[6pt]
\frac{d}{dt}\langle \hat{n}_j \rangle \;=\;& 
-2\,\gamma_j\,\langle \hat{n}_j\rangle
- iJ\,(\delta_{1j} - \delta_{2j})
   \Bigl(\langle \hat{a}_1^{\dagger}\hat{a}_2 \rangle 
         - \langle \hat{a}_2^{\dagger}\hat{a}_1\rangle \Bigr)
+ iF(t)\,\Bigl(\langle \hat{a}_j\rangle 
         - \langle \hat{a}_j^{\dagger}\rangle \Bigr),
\label{eq:n_j}
\\[6pt]
\frac{d}{dt}\langle \hat{a}_j^2\rangle \;=\;& 
-2\bigl(\gamma_j + i\Delta_j\bigr)\,\langle \hat{a}_j^2\rangle
-2\,i\,J\,\langle \hat{a}_1\hat{a}_2\rangle
-2\,i\,F(t)\,\langle \hat{a}_j\rangle 
\nonumber \\
& \quad 
- i\,U_j \Bigl( 
   \langle \hat{a}_j^2\rangle 
   + 6\,\langle \hat{n}_j\rangle\,\langle \hat{a}_j^2\rangle 
   - 4\,\langle \hat{a}_j^{\dagger}\rangle\,\langle \hat{a}_j\rangle^3 
\Bigr),
\label{eq:a_j2}
\\[6pt]
\frac{d}{dt}\langle \hat{a}_1^{\dagger}\hat{a}_2\rangle \;=\;& 
i\,(\Delta_1 - \Delta_2)\,\langle \hat{a}_1^{\dagger}\hat{a}_2\rangle 
- (\gamma_1+\gamma_2)\,\langle \hat{a}_1^{\dagger}\hat{a}_2\rangle 
\nonumber \\
& \quad 
- i\,J\,\Bigl(\langle \hat{n}_1\rangle - \langle \hat{n}_2\rangle\Bigr) 
+ i\,F(t)\,\Bigl(\langle \hat{a}_2\rangle - \langle \hat{a}_1^{\dagger}\rangle \Bigr) 
\nonumber \\
& \quad 
+ i\,U_1 \Bigl( 
   2\,\langle \hat{n}_1\rangle\,\langle \hat{a}_1^{\dagger}\hat{a}_2 \rangle 
   + \langle \hat{a}_1^{\dagger2}\rangle\,\langle \hat{a}_1\hat{a}_2 \rangle 
   - 2\,\langle \hat{a}_1^{\dagger2}\rangle\,\langle \hat{a}_1 \rangle\,\langle \hat{a}_2\rangle
\Bigr) 
\nonumber \\
& \quad 
- i\,U_2 \Bigl( 
   2\,\langle \hat{n}_2\rangle\,\langle \hat{a}_1^{\dagger}\hat{a}_2 \rangle 
   + \langle \hat{a}_2^2\rangle\,\langle \hat{a}_1^{\dagger}\hat{a}_2^{\dagger} \rangle 
   - 2\,\langle \hat{a}_1^{\dagger} \rangle\,\langle \hat{a}_2^{\dagger}\rangle\,\langle \hat{a}_2\rangle^2 
\Bigr),
\label{eq:a1daggera2}
\\[6pt]
\frac{d}{dt}\langle \hat{a}_1\hat{a}_2\rangle \;=\;& 
-\sum_{j=1,2}\bigl(\gamma_j + i\Delta_j \bigr)\,\langle \hat{a}_1\hat{a}_2\rangle 
- i \sum_{j=1,2}\Bigl(J\,\langle \hat{a}_j^2\rangle + F(t)\,\langle \hat{a}_j\rangle\Bigr) 
\nonumber \\
& \quad 
- i\,U_1 \Bigl( 
   2\,\langle \hat{n}_1\rangle\,\langle \hat{a}_1\hat{a}_2 \rangle 
   + \langle \hat{a}_1^{\dagger}\hat{a}_2\rangle\,\langle \hat{a}_1^2\rangle 
   - 2\,\langle \hat{a}_1^{\dagger}\rangle\,\langle \hat{a}_1^2\rangle\,\langle \hat{a}_2\rangle 
\Bigr) 
\nonumber \\
& \quad 
- i\,U_2 \Bigl( 
   2\,\langle \hat{n}_2\rangle\,\langle \hat{a}_1\hat{a}_2 \rangle 
   + \langle \hat{a}_2^{\dagger}\hat{a}_1\rangle\,\langle \hat{a}_2^2\rangle 
   - 2\,\langle \hat{a}_2^{\dagger}\rangle\,\langle \hat{a}_1\rangle\,\langle \hat{a}_2^2\rangle 
\Bigr),
\label{eq:a1a2}
\end{align}
where \(\delta_{ij}\) is the Kronecker delta. In the main text, we assume \(\Delta_1 = \Delta_2\) and \(\gamma_1 = \gamma_2\) for simplicity. 

\begin{figure}[!ht]
\centering
\includegraphics[keepaspectratio, width = 0.47\textwidth]{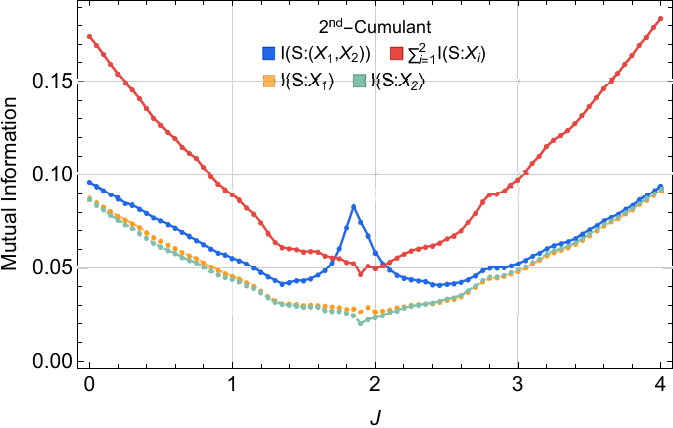}
\caption{Classical mutual informations for the 2\textsuperscript{nd}-order cumulant description. This is to be contrasted with Fig. \ref{fig:prototype_result} (left) to reveal how second-order description captures partial but nontrivial correlation effects correcting mean-field approximation. We set $\Delta = -2, F = 0.5, \gamma = 0.5, U_1 = 0.2, U_2 = 2U_1$ to represent a dynamical regime with non-negligible correlations, motivating the use of second-order cumulants description.}
\end{figure}

Note that these second-order equations provide an approximation to go beyond a strict mean-field approximation without the full computational cost of higher-order correlation. By discarding third- and fourth-order cumulants, we retain the information that captures pairwise correlations, which often dominate many relevant dynamics, while avoiding an intractable explosion in the number of degrees of freedom.
\section*{References}
\bibliography{main}

\end{document}